\documentclass[twoside]{article}
\usepackage{PRIMEarxiv}
\usepackage{authblk}
\pdfoutput=1

\usepackage[utf8]{inputenc} 
\usepackage[T1]{fontenc}    
\usepackage[hidelinks]{hyperref}       
\usepackage{url}            
\usepackage{amsfonts}       
\usepackage{booktabs}       
\usepackage{nicefrac}       
\usepackage{microtype}      
\usepackage{lipsum}
\usepackage{fancyhdr}       
\usepackage{graphicx}       
\graphicspath{{media/}}     

\usepackage{xcolor}
\usepackage{amsmath,amssymb}

\usepackage{appendix}
\usepackage{cleveref}

\usepackage{algorithm}
\usepackage{algorithmic}

\newcommand{\E}{\boldsymbol{E}}
\newcommand{\boldy}[1]{\boldsymbol{#1}}
\newcommand{\Real}{\text{Re}}

\title{Modelling floral and arthropod electrostatics using a two-domain AAA-least squares algorithm}

\author[1]{Samuel J. Harris} 
\author[2*]{Ryan A. Palmer}
\author[1]{N. R. McDonald}

{\affil[1]{Department of Mathematics, University College London, London, WC1E 6BT, UK}
\affil[2]{School of Engineering Mathematics and Technology, University of Bristol, University Walk, Bristol, BS8 1TW, UK}
\affil[*]{Corresponding author: ryan.palmer@bristol.ac.uk}}

\begin{document}
\maketitle

\begin{abstract}
Relationships between plants and insects vitally underpin the health of global ecosystems and food production. 
Through coevolution, insects have acquired a variety of senses in response to the emergence of floral cues such as scent, colour, and shape. 
Therefore, the recent discovery of electroreception among terrestrial arthropods motivates the investigation of floral electrics as part of their wider sensory ecology.
We examine how a flower's morphology and material properties produce and propagate detectable, ecologically relevant electrical signals in several biologically inspired scenarios. As the electrical field both interior and exterior to the flower must be solved for, we develop an extension of the two-dimensional AAA-least squares algorithm for solving such two-domain electrostatics problems. It is found that the electrical signals produced by the plant can reveal information to the insect about the flower shape, available pollen, and the presence of other nearby arthropods. These results show good qualitative agreement with an equivalent three-dimensional scenario, computed using finite element methods. The extension of the AAA-least squares algorithm to two-domain problems provides a fast and accurate method for modelling electrostatic problems, with possible further application in fluid dynamics and magnetostatics. Biologically, our results highlight the significant role floral electrics may play in plant-pollinator and predator-prey relationships, unveiling previously unstudied facets of these key relationships.
\end{abstract}


\section{Introduction}\label{sec:intro}
The advent of spring brings a hive of activity among flora and fauna alike.
As Earth's seasonal cycle runs its course, the blossoming of flowers and emergence of pollinators indicate the arrival of spring throughout the world's temperate climates.
The annual appearance
of new life is key to the health of ecosystems around the world that critically depend upon the interactions between flowers and pollinators \cite{memmott1999structure, bascompte2007plant}.
Pollination underpins the reproduction of many plants and with it the production of important food sources for humans and animals \cite{klein2007importance, rader2016non, khalifa2021overview}.
Consequently, understanding the relationship between plants and pollinators is of utmost importance, especially in light of the increasingly disruptive impact of climate change on biodiversity, food security, and agriculture around the world \cite{memmott2007global, pudasaini2015effect, mbow2020food}. 

\subsection{Terrestrial electroreception}
The relatively recent discovery of electroreception \cite{clarke2013detection, morley2018electric} among terrestrial arthropods, such as bees \cite{greggers2013reception, amador2017honey, clarke2017bee}, spiders \cite{morley2018electric}, hoverflies \cite{khan2021electric}, and ticks, \cite{england2023static} has unveiled a new aspect to plant-arthropod relationships.
This sensory capability of an arthropod to detect and use electrical fields raises questions about not only the role of electrical interactions in nature \cite{hunting2021tree, hunting2022synthetic, hunting2022observed} but also the wider electrical ecology of terrestrial environments at large \cite{hunting2021challenges, hunting2021atmospheric, england2021ecology}.

There have been several studies into the electrical sensory abilities of arthropods. Primarily, these studies have focused on some of the remarkable behavioural implications in communication \cite{greggers2013reception}, foraging \cite{amador2017honey, khan2021electric}, dispersal \cite{morley2018electric}, parasitism \cite{england2023static}, and pollination \cite{clarke2017bee}.
Much attention has also been paid to the mechanisms by which arthropods may sense electrical fields.
Mechanosensory hairs are currently the putative sensors by which arthropods sense electrical fields \cite{sutton2016mechanosensory}.
The current literature focuses on the topics of hair response to electrical stimuli \cite{koh2020bumblebee, rpalmer2021analysis}, collective sensor dynamics \cite{palmer2022mechanics}, and possible uses of electroreception \cite{palmer2023passive, palmer2023analysis}.

Furthermore, the coevolution of plants and arthropods over many millennia has led to synergistic behaviours and relationships \cite{ patiny2011evolution, balamurali2015senses, dar2017pollination}.
This includes diversification in flower size, scent, colour, and shape and the emergence of different sensory perceptions and receptors in arthropods. Thus, plant-pollinator coevolution and the behavioural and mechanical evidence of electroreception motivate investigation of floral electrics as part of the rich ecology of plants and arthropods.

\subsection{Floral electrostatics}
Several papers have considered the electrostatics of plants \cite{hunting2022synthetic,moyroud2017physics, montgomery2021bumblebee,  nanda2022study, molina2023electromagnetic}.
Yet, further research into the generation and characteristics of floral electrical fields is required. Such studies may reveal new, previously unknown facets of how plants and arthropods exchange information via electrical fields. 

It has been shown that bees are predominantly positively charged to $ \mathcal{O}(10) - \mathcal{O}(100) \ pC$ \cite{clarke2013detection}.
When approaching an uncharged flower, the pollinator's presence is hypothesised to polarise the flower, producing an electric field \cite{clarke2017bee}. The strength and characteristics of this floral electric field depend on its shape, proximity to the arthropod, and propensity to polarise (i.e. relative permittivity, the scale by which an electric field between the charges decreases relative to a vacuum). 
On this last point, the petals and pollen of flowers are broadly considered to be dielectric since, at their surface, they consist of waxy material with little conductivity \cite{nanda2022study, kundu2014broadband}.

Therefore, the physics of the studied system is that of two-domain dielectric polarisation. In particular, both the electrical field exterior to the flower and that induced within the (dielectric) flower structure must be considered, with suitable matching conditions imposed on the flower boundary, for example, continuity of the electric potential. Mathematically, this is an added level of complexity from the simple one-domain problem and from multiply connected domains which still consider only one region punctured by multiple ``holes.'' In the two-domain scenario, the coupled system of both interior and exterior governing equations must be solved simultaneously. By focusing on the interaction of individual flowers and their local electrical environment, we seek to understand how a flower's morphology and material properties produce and propagate detectable, ecologically relevant electrical signals.

\subsection{The AAA-least squares algorithm}
We consider several biologically inspired scenarios in two (2D) and three dimensions (3D). In 2D, we develop a new method for solving such two-domain electrostatic problems by adapting the combined adaptive Antoulas-Anderson (AAA) and least-squares (LS) algorithm (AAA-LS) developed by Trefethen and colleagues; see, e.g., \cite{trefethen2020numerical,costa2023aaa,nakatsukasa2023first}. Mathematically, the problem is to find the harmonic electric potential $V$ interior and exterior to the flower, subject to prescribed conditions at the flower-air interface and given behaviour in the far-field. This potential $V$ can be written as the real part of an analytic function $F(z)$ approximated by rational functions: a polynomial plus a sum of singular terms involving poles clustered near corners/cusps on the flower boundary. While these poles can be placed a priori close to the corners as in the open source code the ``lightning Laplace solver'' (see \cite{gopal2019solving,trefethen2020numerical}), the AAA algorithm (pronounced triple-A) can instead be used to adapt the singularities to the solution directly \cite{costa2023aaa}. The function $F$ is then found using a linear LS algorithm. 

This combined AAA-LS method has been applied to a variety of  problems in mathematics, physics, chemistry, and biology \cite{nakatsukasa2023first,harris2023penguin,xue2023rational,kehry2023robust} and extended \cite{gopal2019solving,nakatsukasa2023first} to incorporate multiply connected domains \cite{trefethen2020numerical,costa2023aaa} and other governing equations such as the Stokes \cite{brubeck2022lightning}, Poisson \cite{harris2023penguin}, and Helmholtz equations \cite{gopal2019new}. The work of \cite{costa2024modelling} gives an example of the AAA algorithm being used to model the internal and external magnetic fields of a physical object by considering each field separately. However, to the best of our knowledge, there is currently no published work on extending the AAA-LS method to general, inhomogeneous, two-domain interface problems, as also noted in \cite{costa2024modelling}. Such an extension is realised in this paper by utilising the two boundary conditions (of Dirichlet and Neumann types) to solve for the two unknown interior and exterior potentials $V_1$ and $V_2$. The results of this two-domain AAA-LS method are compared against known solutions for circular and elliptical geometries (see section~\ref{sec:aaals}) and found to achieve accuracy comparable to that typically found in the one-domain algorithm \cite{costa2023aaa}.

In 3D, finite element methods (FEM) make up the most common approach to numerically solve the electrostatic flower-arthropod problem. 
While FEM can be reasonably accurate, they struggle to capture details across multiple-scales and can become computationally time-consuming with increased fidelity. Another downside is the need to create new geometries and meshes for each new scenario or morphology one may wish to examine. Therefore, the FEM approach is less suited to a broad scenario analysis of floral geometry. Henceforth, the 2D problem is primarily considered throughout this work and shown to be a reasonable indicator of the full 3D behaviour. 

There are two key purposes of this work. First, it presents a new extension of the existing AAA-LS algorithm to two-domain geometries. This method finds a rational approximation of a harmonic function both interior and exterior to some closed curve, with boundary conditions coupling the two regions. Second, it gives a large-scale, quantitative analysis of the electrostatic interaction between flowers and arthropods and of whether certain floral features, such as flower shape and available pollen, produce distinct electrical signatures detectable to arthropods. While the electrostatics of both flowers and arthropods have been individually studied, the floral-arthropod study performed here with a detailed description of the flower geometry is new. Both of these research findings complement each other: the floral-arthropod interaction provides a relevant real-world scenario in which to apply the two-domain AAA-LS method to. Further, the speed and accuracy of the AAA-LS method are well-suited to the broad scenario analysis of floral-arthropod electrostatics undertaken in this work.

The paper continues with an overview of the model setup in section~\ref{sec:model}. Section~\ref{sec:aaals} provides the technical details for the adaptation of the AAA-LS algorithm to the new two-domain setting. The proceeding two sections present a systematic analysis of floral electrical fields, beginning in section~\ref{sec:results} with (i) flowers in uniform electrical fields, (ii) polarisation in the presence of a pollinator, (iii) the electrostatic effect of pollen, (iv) multiple foragers, and (v) predator-prey interactions. Then, a 3D comparison of the flower-pollinator results are shown in section~\ref{sec:3D}.
Finally, discussion and conclusion of the results is presented in section~\ref{sec:conc}, including reference to the biological relevance of the results and their limitations and possible avenues for future work.

\begin{figure}[t]
    \centering
\includegraphics[width=0.75\linewidth]{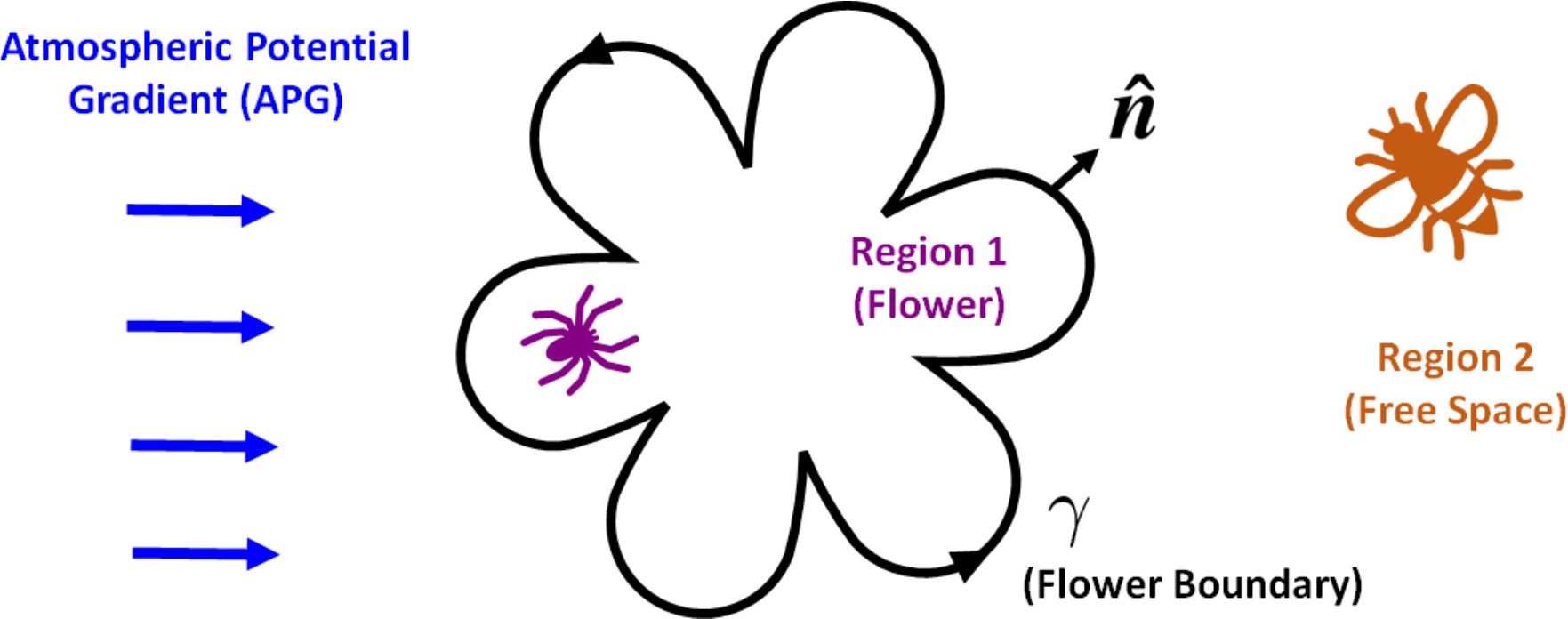}
    \caption{Diagram of the arthropod and uniform field problem. Three types of electrical source/sink are shown: (i) an external planar field (e.g., the Earth's atmospheric potential gradient), (ii) a point charge outside of the flower boundary, and (iii) a point charge inside the flower boundary.}
    \label{fig:model}
\end{figure}

\section{Model setup}
\label{sec:model}
Consider a flower which is treated as a 2D dielectric; see Figure \ref{fig:model}. The shape of the flower is traced out by some Jordan curve $\gamma$ orientated such that the unit normal vector $\hat{\boldsymbol{n}}$ is outward pointing. Let the interior of the flower be labelled as region 1 and the exterior as region 2.

An electric field $\E$ is present in both regions 1 and 2 due to the dielectric nature of the flower. This field can be written as the gradient of some scalar potential $\E=-\boldsymbol{\nabla}V$ which satisfies the Poisson equation $\nabla^2V = -\rho/\epsilon$ where $\rho$ is the charge density and $\epsilon=\epsilon_0(1+\chi_e)=\epsilon_0\epsilon_r$ is the absolute permittivity. Here, $\epsilon_0\approx8.85\times10^{-12}\text{ Fm}^{-1}$ is the permittivity of free space, $\chi_e$ the electric susceptibility and $\epsilon_r=1+\chi_e$ the relative permittivity. In the exterior (free space) region, it holds that $\epsilon_r=1$ and so $\epsilon_2=\epsilon_0$, whereas in the interior flower region $\epsilon_r = 10$ to $100$ and so $\epsilon_1= 10\epsilon_0$ to $100\epsilon_0$.

Dimensionless quantities are introduced via the following scalings, where henceforth starred variables are dimensional. Lengths are scaled as $\boldy{x}^\ast = L\boldy{x}$, where $L$ is the length scale of the flower; it follows that $\nabla^\ast = (1/L)\nabla$. There are three possible generators of the electric field $\E$, which will determine the scaling of the potential $V$:

\begin{enumerate}
    \item[(a)] A uniform electric field such as the atmospheric potential gradient (APG).
    \item[(b)] Arthropods exterior (e.g. bees) and interior (e.g. spiders) to the flower.
    \item[(c)] A combination of (a) and (b). This is illustrated in Figure \ref{fig:model}.
\end{enumerate}
The governing equations and boundary conditions are identical in all three problems and independent of the scaling of $V$. 
By Gauss's law, the divergence of the electric field is everywhere zero except at point charges (arthropods) where the derivatives are not defined. Excluding these singularities, the potential $V$ satisfies the Laplace equation in both interior and exterior regions. The system of dimensionless equations is thus
\begin{gather}
    \nabla^2V_1=0\;\;\;\text{in 1,}\label{Meq:ilap}\\
    \nabla^2V_2=0\;\;\;\text{in 2,}\label{Meq:elap}\\
    V_1=V_2\;\;\;\text{on }\gamma, \label{Meq:dBC}\\
    \frac{\partial V_1}{\partial n}=\tilde{\epsilon}\frac{\partial V_2}{\partial n}\;\;\;\text{on }\gamma,\label{Meq:nBC}
\end{gather}
where $\tilde{\epsilon}=\epsilon_2/\epsilon_1$ is the ratio of the absolute permittivities. Boundary conditions~\eqref{Meq:dBC} and~\eqref{Meq:nBC} are consequences of the electrical potential's continuity across the boundary and of Gauss's law, respectively \cite{griffiths2005introduction}. The far-field condition on the potential $V_2$ depends upon the generator of $\E$. 

\subsection{Uniform electric field}\label{subsec:uniform}
Let a uniform background electric field be present with strength $E_\infty$ far from the flower. The interior and exterior potentials $V_1$ and $V_2$, respectively, are scaled by $LE_\infty$ as $V^\ast = LE_\infty V$. The far-field condition is thus
\begin{equation}
    V_2\rightarrow  -x\;\;\;\text{as }r\rightarrow\infty. \label{Meq:ufar}
\end{equation}
where $x$ is the horizontal direction and $r$ is the radial direction. When the flower boundary $\gamma$ is a circle or an ellipse, an exact solution to the system \eqref{Meq:ilap}-\eqref{Meq:ufar} can be found; these solutions are given in Appendix \ref{appendix1}.
In general, the uniform electrical field may possess an ``out of plane'' component, relative to the flower. In this instance we only consider the component that lies in-plane with the flower. Additionally, since the orientation of the flower within the model is arbitrary, the $x$-axis can be chosen to align with the uniform field.

\subsection{Bees and spiders}\label{subsec:bee}
Consider an arthropod with associated electric charge $\lambda$. The creature is centered at the point $z_1^\ast$ which is typically of $\mathcal{O}(L)$ and can be either inside (spider) or outside (bee) the flower.
It is assumed the arthropod is not on the flower boundary and is sufficiently small to be considered as a point charge. 
In a 2D system, the electric potential of a point charge in free space 
is known as
\begin{equation}\label{epotfree}
    V^\ast = -\frac{\lambda}{2\pi\epsilon}\log|z^\ast-z_1^\ast|=-Q\log|z^\ast-z_1^\ast|,
\end{equation}
where the absolute permittivity is $\epsilon=\epsilon_1|\epsilon_2$ for $z_1\in 1|2$. The potential $V$ is scaled by $Q$ as $V^\ast = QV$, and so the far-field condition is
\begin{equation}
    V_2\rightarrow -\log|z-z_1|-\log|L|\;\;\;\text{as }r\rightarrow\infty.\label{Meq:spfar}
\end{equation}
Exact solutions to the system \eqref{Meq:ilap}-\eqref{Meq:nBC}, \eqref{Meq:spfar} can be found for a circular flower boundary; these are given in Appendix \ref{appendix2}.

\subsection{Swarming arthropods in the electric field}\label{subsec:combined}
The problems of sections \ref{subsec:uniform} and \ref{subsec:bee} can be combined to model how an arthropod of electric charge $\lambda_1$ interacts with a uniform electric field of far-field strength $E_\infty$. The potential $V$ is again scaled as $V^\ast = LE_\infty V$, giving the far-field condition
\begin{equation}
    V_2\rightarrow -x - A_1\log|z-z_1|-B_1\;\;\;\text{as }r\rightarrow\infty, \label{Meq:cfar}
\end{equation}
where the dimensionless parameter $A_1=Q_1/(LE_\infty)$ and the constant $B_1=A_1\log|L|$ have been introduced. Including multiple arthropods (a swarm) can be achieved by including additional $A_j\log|z-z_j|+B_j$ terms on the right-hand side of \eqref{Meq:cfar}, where arthropod $j$ centered at the point $z_j^\ast=Lz_j$ has charge $\lambda_j$. If there is no background electric field, we exclude the $-x$ term in \eqref{Meq:cfar} and create the new dimensionless parameters $A_j = Q_j/Q_1$. The exact solution of \eqref{Meq:ilap}-\eqref{Meq:nBC}, \eqref{Meq:cfar} for a circular flower is just a linear combination of the solutions in Appendices \ref{appendix1} and \ref{appendix2}.

\section{AAA-LS numerical method}
\label{sec:aaals}
All three variants of the flower problem in section \ref{sec:model} can be solved numerically using an extension of the AAA-LS method developed by Trefethen and colleagues \cite{trefethen2020numerical,costa2023aaa,nakatsukasa2023first}. This numerical approach combines the adaptive Antoulas Anderson (AAA) algorithm, used to find a rational approximation of the boundary data, followed by a least squares (LS) fit to this data to approximate the solution of a Laplace problem in terms of a polynomial and a finite sum of rational functions. The algorithm is fast and accurate, running in a fraction of a second on a standard laptop and converging root-exponentially with respect to the number of poles generated by the AAA algorithm \cite{costa2023aaa,nakatsukasa2018aaa}. Further, it can handle non smooth domain shapes such as those with corners and/or cusp singularities.

The original algorithm \cite{costa2023aaa} solves the one-domain Dirichlet problem
\begin{equation}
    \nabla^2\phi=0\;\;\text{in }D, \;\;\;\;\;\;\;\phi=h(z)\;\;\text{on }\gamma,
\end{equation}
where $D$ is either the interior or exterior domain to a Jordan curve $\gamma$ and $h$ is a given function. The algorithm can also be used for problems with Neumann boundary condition $\phi_n=0$ on $\gamma$. The potential $\phi$ can be expressed as the real part of an analytic function $F(z)$ in $D$ which, in turn, is approximated by 
\begin{equation}\label{eq:onedomainF}
    \phi = \Real[F(z)] = \Real\left( \ \sum_{k=1}^N a_k (z-z_c)^k + \sum_{k=1}^M \frac{b_k}{z-p_k}\right),
\end{equation}
where $N$ is the series truncation (for numerical purposes). The first sum is known as the smooth (Runge) part and the second as the singular (Newman) part. The form \eqref{eq:onedomainF} assumes that $D$ is the interior domain; the powers of $(z-z_c)$ are negative in the Runge part when considering the exterior domain. The point $z_c$ is located in the interior domain with the choice $z_c=0$ made in this work. The AAA algorithm uses the boundary condition for $\phi$ to find the $M$ poles $p_k$ which lie in the region $\Omega = \mathbb{C}\backslash D$. The complex coefficients $a_k$ and $b_k$ are then found using an LS algorithm applied to a set of points on the boundary $\gamma$: constructing a matrix of basis vectors $A$ and the vector $H=h(z_b)$, the vector of coefficients $c=[a_k; b_k]$ can be found by the backslash operation $c=H\backslash A$. Construction of the Vandermonde matrix $A$ can be coupled with an Arnoldi orthogonalisation for added stability; see \cite{brubeck2021vandermonde}. However, we do not perform this additional step in the present work as no such instabilities were evident. The one-domain AAA-LS algorithm is summarised in Algorithm \ref{alg:onedomainAAA}.

\begin{algorithm}
\caption{One-domain AAA-LS algorithm}
\label{alg:onedomainAAA}
\begin{algorithmic}[1]
\STATE{Input boundary data $\gamma: z = z_b$, interior point $z_c$, series truncation $N$ and boundary condition $\phi=H=h(z_b)$ on $\gamma$.}
\STATE{Run AAA algorithm to find suitable exterior poles $p_k$ for given $H$.}
\STATE{Create matrix $A$ of basis vectors $(z_b-z_c)^k$ and $1/(z_b-p_k)$.}
\STATE{Run LS algorithm to find vector of coefficients $c = [a_k;b_k] = H\backslash A$.}
\STATE{Form $\phi$ from \eqref{eq:onedomainF}.}
\end{algorithmic}
\end{algorithm}

The difficulty in adapting the AAA-LS method to a two-domain scenario is that there is no ``given'' function $h$ for the Dirichlet (or Neumann) boundary condition. Instead, there is usually a pair of boundary conditions imposing either continuity or jumps in the unknowns $V_1$ and $V_2$ and their normal derivatives. This poses two challenges: first, in adapting the AAA algorithm, which normally uses a known function $h$ to identify interior/exterior poles; and second, in adapting the LS algorithm, which uses $h$ in the backslash operation to find the unknown coefficients $a_k$, $b_k$. The proposed remedy is to consider the combined quantity $V_1-V_2$ with known functions to be used in the AAA and LS algorithms arising from the far-field conditions.

Since both the interior and exterior potentials $V_1$ and $V_2$ are harmonic in \eqref{Meq:ilap}, \eqref{Meq:elap},  each can be written as the real part of some analytic function plus some correction term to account for the far-field condition. Consider the general case of a system of $J$ arthropods in a uniform electric field, where arthropod $j$ is located at  $z=z_j$. The potentials $V_1$ and $V_2$ are expressed as
\begin{align}
    & \begin{aligned}
    V_1 &= -\Real[G_1(z)]+\Real[F_1(z)]\\
    & = -\Real\bigg(\sum_{j=1}^J\Gamma\tilde{\epsilon}A_j\log(z-z_j)\bigg)+\Real\bigg(\sum_{k=1}^{N_1} a_k (z-z_c)^k + \sum_{k=1}^{M_1}\frac{b_k}{z-p_k}\bigg),  
    \end{aligned} \label{eq:V1aaa} \\ 
    & \begin{aligned}
    V_2 & = -\Real[G_2(z)] + \Real[F_2(z)]\\ 
    &= -\Real\bigg(z + \sum_{j=1}^JA_j\log(z-z_j)\bigg) + \Real\bigg(\sum_{k=1}^{N_2} c_k (z-z_c)^{-k}+\sum_{k=1}^{M_2}\frac{d_k}{z-q_k}\bigg),
    \end{aligned}\label{eq:V2aaa}
\end{align}
where $\Gamma=1|0$ for $z_j\in 1|2$, $z_c=0$ is the centre of the flower and $N_1$ and $N_2$ are series truncations; the values $N_1=N_2=20$ are typically used in this work. The dimensionless parameters $A_j$ are given in section \ref{subsec:combined} and if there is no uniform electric field, the $z$ term in \eqref{eq:V2aaa} can be excluded.

Now, consider the Dirichlet boundary condition \eqref{Meq:dBC} expressed as $V_1-V_2=0$ on $\gamma$. Substituting \eqref{eq:V1aaa} and \eqref{eq:V2aaa} into \eqref{Meq:dBC} and rearranging gives
\begin{equation}\label{eq:dBCaaa}
    \Real[F_1(z)]-\Real[F_2(z)] = -\Real[z + A_j(1-\Gamma\tilde{\epsilon})\log(z-z_j)] = -\Real[H_1(z)],
\end{equation}
where the summation sign has been dropped for brevity. The Neumann boundary condition \eqref{Meq:nBC} can be expressed similarly. Note that $\partial V/\partial n = \boldsymbol{\hat{n}\cdot\nabla}V = \Real[n\overline{\nabla}V]$, where $n=n_x+in_y$ and $\nabla$ are complex representations of the normal vector to $\gamma$ and the gradient operator. Further, equation (5) from \cite{trefethen2018series} gives that $\overline{\nabla}[\Real[F(z)]] = F'(z)$. Therefore, the Neumann boundary condition becomes
\begin{equation}\label{eq:nBCaaa}
    \Real[nF_1'(z)]-\tilde{\epsilon}\Real[nF_2'(z)] = -\Real[\tilde{\epsilon}n + A_j\tilde{\epsilon}(1-\Gamma)n/(z-z_j)] = -\Real[H_2(z)].
\end{equation}
The functions $H_1$ and $H_2$ are known and can be evaluated for some given boundary data $z_b$, therefore it is these functions that are to be used in the AAA and LS algorithms.

First, the collection of $M_1$ poles $p_k$ in region 1 for $V_1$ and $M_2$ poles $q_k$ in region 2 for $V_2$ are found using the AAA algorithm. This is achieved by finding interior and exterior poles relevant to the function $-\Real[H_1(z)]$ and labelling these as $p_{dk}$ and $q_{dk}$, respectively, to signify these are the poles relating to the Dirichlet boundary condition. Similarly, the AAA algorithm is used again to find the poles relevant to the function $-\Real[H_2(z)]$ with these labelled as $p_{nk}$ and $q_{nk}$. Combining the two sets then gives all the required poles $p_k = [p_{dk} \;\;p_{nk}]$ and $q_k = [q_{dk} \;\;q_{nk}]$. There may be the occurrence of ``over counting'': singularities resulting from $V_1$ also appear (unnecessarily) in the equation for $V_2$ and vice versa. Similarly, ``under counting'' may occur if a pole relevant to only one potential is not relevant to the combined quantity $V_1-V_2$. These occurrences are inconsequential here as any exponential clustering of poles near corner singularities \cite{gopal2019solving} is suitable for use in \eqref{eq:V1aaa} and \eqref{eq:V2aaa}.

Finally, an LS method similar to the one-domain problem is  used to evaluate expressions \eqref{eq:dBCaaa} and \eqref{eq:nBCaaa}. However, the matrices and vectors are now ``twice as large'' to account for the two potentials that are to be found. By creating a matrix of basis vectors $A$ and the vector $H = -\Real[H_1(z); H_2(z)]$, the vector of unknown coefficients $c = [a_k; b_k; -c_k; -d_k]$ can be found again by the backslash operation $c=H\backslash A$. Thus, the potentials $V_1$ \eqref{eq:V1aaa} and $V_2$ \eqref{eq:V2aaa} are found numerically. Using $\nabla[\Real[F(z)]] = \overline{F'(z)}$, the electric field vectors expressed in complex notation as $E=-\nabla V$ are obtained by
\begin{equation}\label{eq:E12def}
    E_i = \overline{G_i'(z)} - \overline{F_i'(z)},
\end{equation}
where $i=1,2$ corresponds to regions 1 and 2 with $G_i,F_i$ given in \eqref{eq:V1aaa} and \eqref{eq:V2aaa}. The two-domain AAA-LS algorithm is summarised in Algorithm \ref{alg:twodomainAAA}.

\begin{algorithm}
\caption{Two-domain AAA-LS algorithm}
\label{alg:twodomainAAA}
\begin{algorithmic}[1]
\STATE{Input boundary data $\gamma: z = z_b$, interior point $z_c$, series truncations $N_1$, $N_2$ and boundary condition functions $H_1$, $H_2$ from \eqref{eq:dBCaaa}, \eqref{eq:nBCaaa}.}
\STATE{Run AAA algorithm to find suitable interior and exterior poles $p_{dk}$, $q_{dk}$ for the function $-\text{Re}[H_1(z)]$.}
\STATE{Run AAA algorithm to find suitable interior and exterior poles $p_{nk}$, $q_{nk}$ for the function $-\text{Re}[H_2(z)]$.}
\STATE{Combine interior and exterior poles as $p_k = [p_{dk}, p_{nk}]$ and $q_k = [q_{dk}, q_{nk}]$.}
\STATE{Create matrix $A$ of basis vectors.}
\STATE{Run LS algorithm to find vector of coefficients $c = [a_k;b_k] = H\backslash A$.}
\STATE{Form $V_1,V_2$ from \eqref{eq:V1aaa}, \eqref{eq:V2aaa} and $E_1,E_2$ from \eqref{eq:E12def}.}
\end{algorithmic}
\end{algorithm}

To test the accuracy of this method, the numerical results from the two-domain AAA-LS algorithm were compared against the exact solutions given in Appendices \ref{appendix1} and \ref{appendix2}. For each of the three flower problems, a circular flower of unit radius was considered; for the problem of the uniform electric field only, ellipses with major and minor axes $a$ and $b$, respectively, were also considered. The interior and exterior potentials were computed on a $100\times100$ grid in the range $x,y\in[-3,3]$ and the relative error between AAA-LS and exact solutions was calculated at gridpoints.

In the problem with no arthropods, the relative error was of $\mathcal{O}(10^{-16})$ for a circular flower, demonstrating excellent agreement between the AAA-LS method and the exact solution. This error was persistently small for near-circular ellipses, for example an ellipse of axes $a=1.3$ and $b=0.7$ gave an error of $\mathcal{O}(10^{-14})$. For more elongated ellipses, the error gradually increased yet remained comparatively small, with an ellipse of axes $a=1.7$ and $b=0.3$ giving an error of $\mathcal{O}(10^{-6})$. Further, results from a FEM model produced in COMSOL 6.1 were also compared against the exact solution and found to have a consistent relative error of $\mathcal{O}(10^{-2})$ for all circle and ellipse examples performed. In problems involving arthropod(s) interior/exterior to a circular flower, the relative errors between AAA-LS and exact solutions were also small, of $\mathcal{O}(10^{-12})$.

Two key results can be drawn from these tests. First, the two-domain AAA-LS method gives highly accurate solutions at the order of magnitude expected from the one-domain algorithm \cite{costa2023aaa}. This helps affirm that the method is performing correctly and gives us the confidence to continue using the algorithm for more complicated geometries. Second, the AAA-LS method consistently outperforms the more standard FEM solver in accuracy by several orders of magnitude. Combined with the algorithm's speed and simplicity, this makes the AAA-LS method a competitive tool to use, as has been noted since its development five years ago \cite{nakatsukasa2023first}. The entirety of the two-domain AAA-LS code used in this work is open access and freely available -- see \cite{flowercode}.

\section{Biologically motivated application: Floral electrical fields} \label{sec:results}
Using the above method, we investigate several scenarios that illustrate possible pollinator-plant interactions.
Our topic of interest is how an uncharged dielectric flower polarises in the presence of some external electrical field, altering the field.
We therefore define the ``perturbation field'' to measure how the flower perturbs the source field as follows:
\begin{align}
    V_P & = V - V_{B}, \\
    \| \mathbf{E}_P \| & = \| \mathbf{E} - \mathbf{E}_B \|,
\end{align}
where the subscript $P$ indicates a perturbation value, $B$ relates to a background electrical field (defined case by case below), and the terms without subscripts are those of the full modelled scenario.
The perturbation field is studied because the external field is often stronger than the flower's polarised field and dominates the results.
It also helps to assess and compare changes in the field strength and structure as the flower ``deforms'' the source electrical field.
From a sensory perspective, the perturbation field assumes that the source (e.g. bee) does not detect itself.
Thus, the perturbation field may be interpreted as the ``floral electrical information'' from an arthropod's perspective.

Finally, it is worth reiterating that the results are nondimensionalised and hold across varying source field strengths, flower sizes and scales, i.e. field strengths are proportional to the source field magnitude. For all upcoming contour plots of the perturbed electrical field magnitude (see for example Figure \ref{fig:uni-flower}), the flower is centred at the origin and has unit petal length. Therefore, the coordinates used in these plots ($x=[-4,4], y=[-4,4]$) can be thought of in terms of petal lengths: the perturbed electrical field at up to four petal lengths away from the flower centre is displayed.
We will discuss the dimensional implications of these results in \cref{sec:conc}.

\subsection{Floral signals in uniform electrical fields}
\begin{figure}
    \centering
    \includegraphics[width=0.7\linewidth,trim=1 1 1 1,clip]{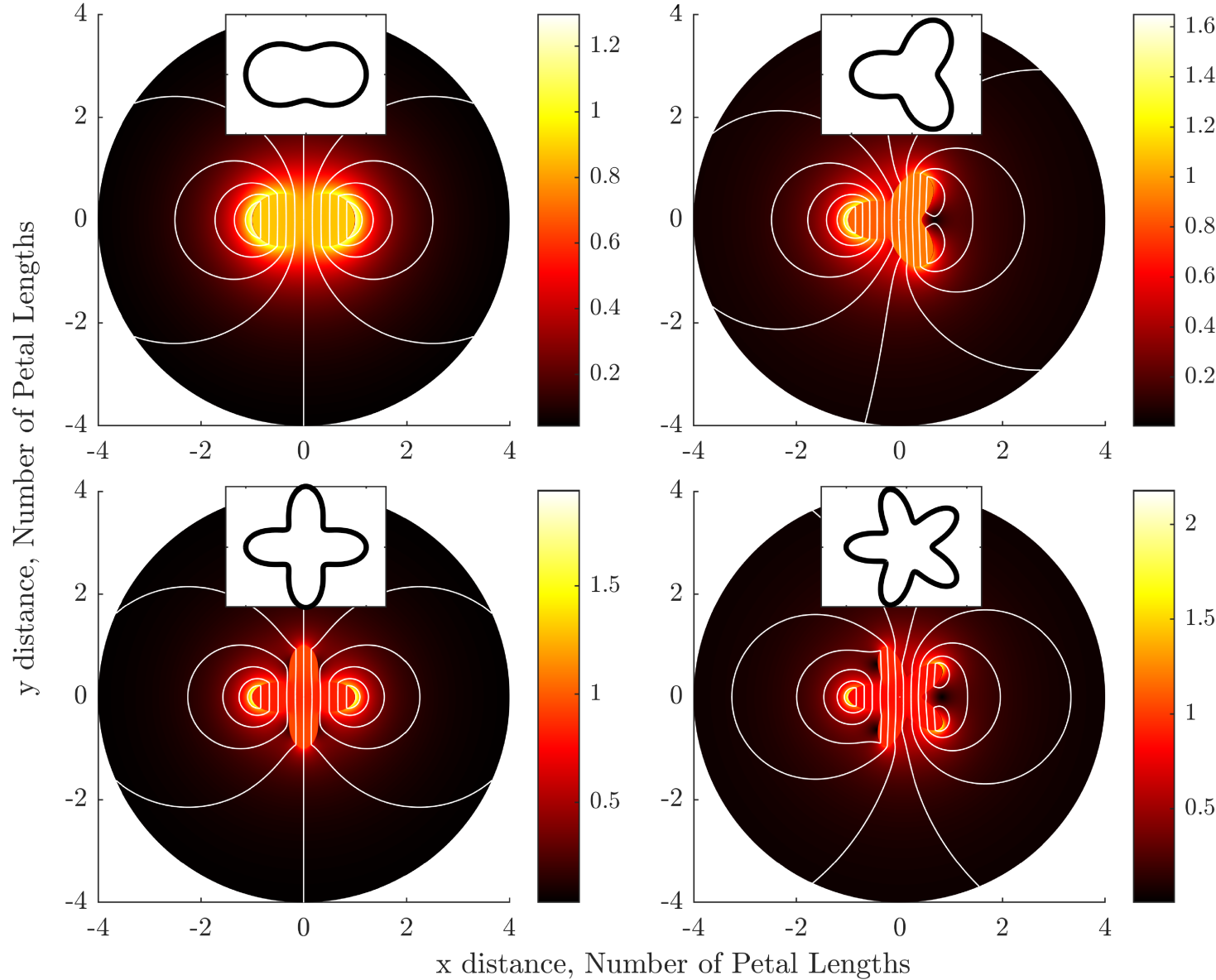}
    \caption{Perturbation electrical field magnitude $\|\mathbf{E}_P\|$ (colourmap) and potential $V_P$ (contours) for an uncharged flower polarising in a uniform electrical field. The flower is centred at the origin and has unit petal length.} The inset plots show the flower geometry. Each shape has the same surface area.
    \label{fig:uni-flower}
    \includegraphics[width=0.8\linewidth]{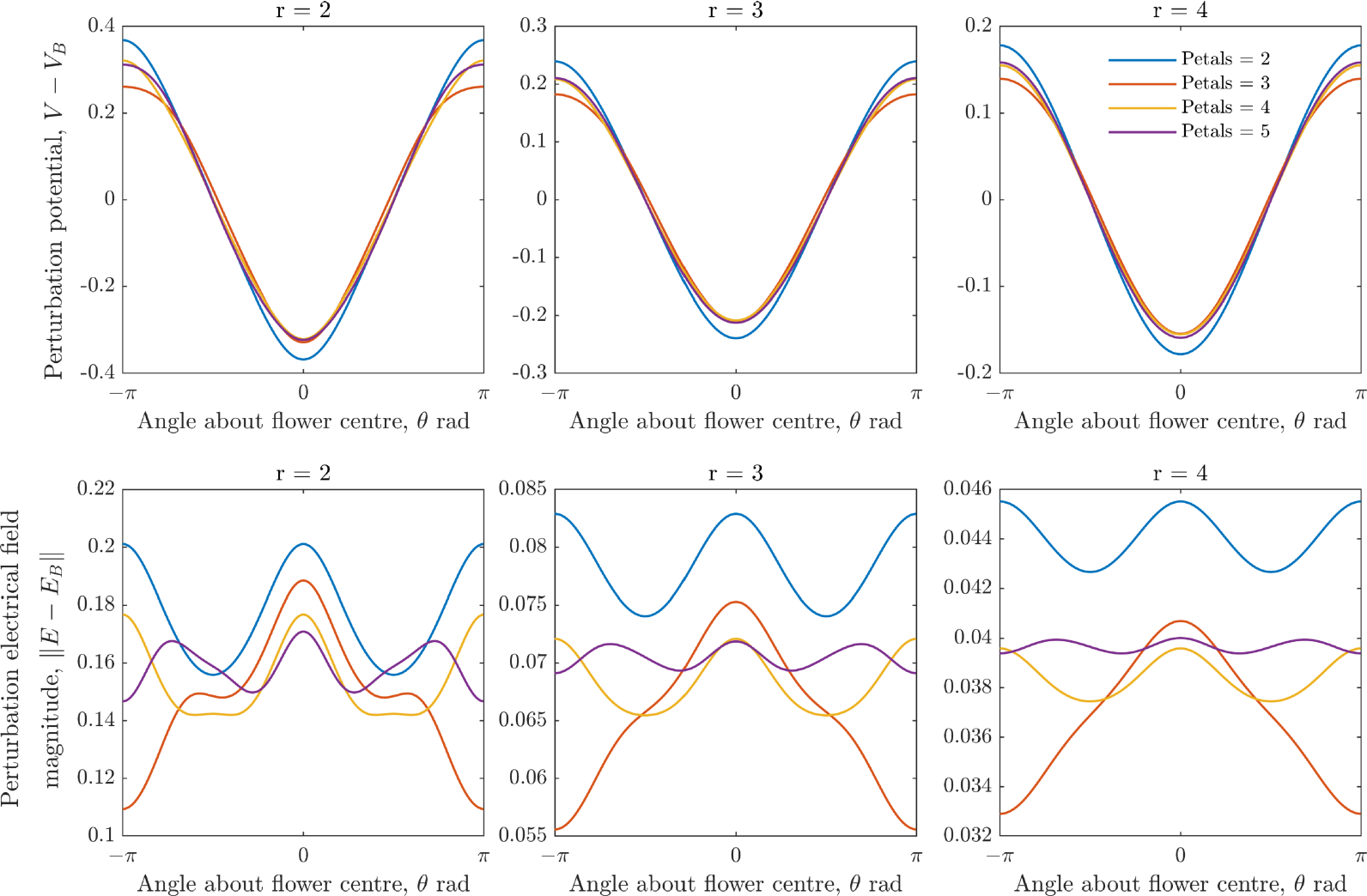}
    \caption{Perturbation electrical potential (top row) and field magnitude (bottom row) of a polarised flower  in a uniform field at increasing radii from the flower centre, $r = 2, 3, 4$.}
    \label{fig:uni-flower-VE}
\end{figure}

Our first example follows the setup presented in \cref{subsec:uniform}.
We consider a flower in an atmospheric potential gradient, which we treat as some locally uniform electrical field such that:
\begin{equation}
    \mathbf{E}_B = -\boldsymbol{x} \ .
\end{equation}
The question of interest is how the presence of a flower deforms this natural background field and provides electrical evidence of its presence at a distance. The perturbation here compares the electrical field with and without the flower present.

Consider four flower shapes ranging from two to five petals.
In each case, the flowers have the same area to ensure no additional area effects are present.
The relative permittivity of the flower is set to 20.

The perturbation electrical field (colourmap) and potential (contours) are shown in Figure \ref{fig:uni-flower} both internally and externally for each flower shape. 
The perturbation electrical potentials show both vertical and horizontal symmetry, for the even-numbered petals, and horizontal symmetry for odd numbered petals, reflecting the floral geometries and their alignment to the source field.
Since the electrical field is the gradient of the potential, larger perturbation field magnitudes are seen for an increasing number of petals due to the stronger variation and therefore the gradients in the flower geometries.
Furthermore, the strongest regions of the perturbed electrical field are found internally in the flower, aligning with the background field. 

We now seek to understand how the flower's polarisation provides information about the flower at a distance.
Figure \ref{fig:uni-flower-VE} presents the values of the perturbation potentials and field magnitudes along circles, centred on the flower, with different radii, $r = 2, 3, 4$, e.g. $1, 2$ and $3$ petal lengths from the flower boundary.
Notably, electroreception is expected to be most potent at $\mathcal{O}(10) cm$ from the flower \cite{clarke2017bee}.
The perturbation potentials are reasonably similar for all flower shapes with a marginal decrease in their magnitudes with increasing petal numbers.
Indeed, the flower geometry leads to little variation in the shape of the potential curves over several petal lengths.
Thus, the change in potential is due to the flower's presence rather than its shape.
Yet, these plots indicate a clear persistence of the perturbation up to three petal lengths or more away from the flower.

\begin{figure}
    \centering
    \includegraphics[width=0.9\linewidth,trim=1 1 1 1,clip]{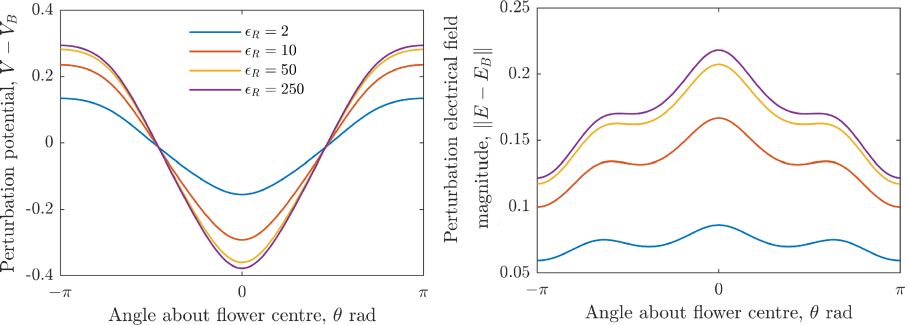}
    \caption{Three-petal flower polarisation in a uniform electrical field at $r = 2$, and varying floral relative permittivity, $\epsilon_R = 2, 10, 50, 250$. Left: Perturbation electrical field magnitude, Right: Perturbation electrical potential.}
    \label{fig:uni-flower-perm}
\end{figure}

Regarding the perturbation fields, Figure \ref{fig:uni-flower-VE} shows the effect of floral geometry.
The petal tip effects at $ r = 2$ are comparatively stronger due to the larger local gradients.
As $r$ increases, the perturbation fields reduce more quickly compared to the potential, as is expected, due to the $1/r$ relationship for the electrical fields compared to the $\log(r)$ dependency of the potentials.
Crucially, the electrical field strengths and variation thereof are the largest contributors to the conveyance of information since these affect the overall force felt by an arthropod's mechanosensors \cite{sutton2016mechanosensory, rpalmer2021analysis, palmer2022mechanics}.
Overall, flower morphology leads to distinct perturbation forms indicating that a flower in a uniform electrical field can display information about its morphology at a distance through electrical fields.
By $r = 4$, the variation is $\mathcal{O}(10^{-3})$, yet considering the acute sensitivity of arthropod mechanosensors and the nondimensionalised results, it is not unreasonable to conclude that these variations will be detectable and determinable by an arthropod (discussed further in \cref{sec:conc}).

\begin{figure}
    \centering
    \includegraphics[width=0.65\linewidth,trim=1 1 1 1,clip]{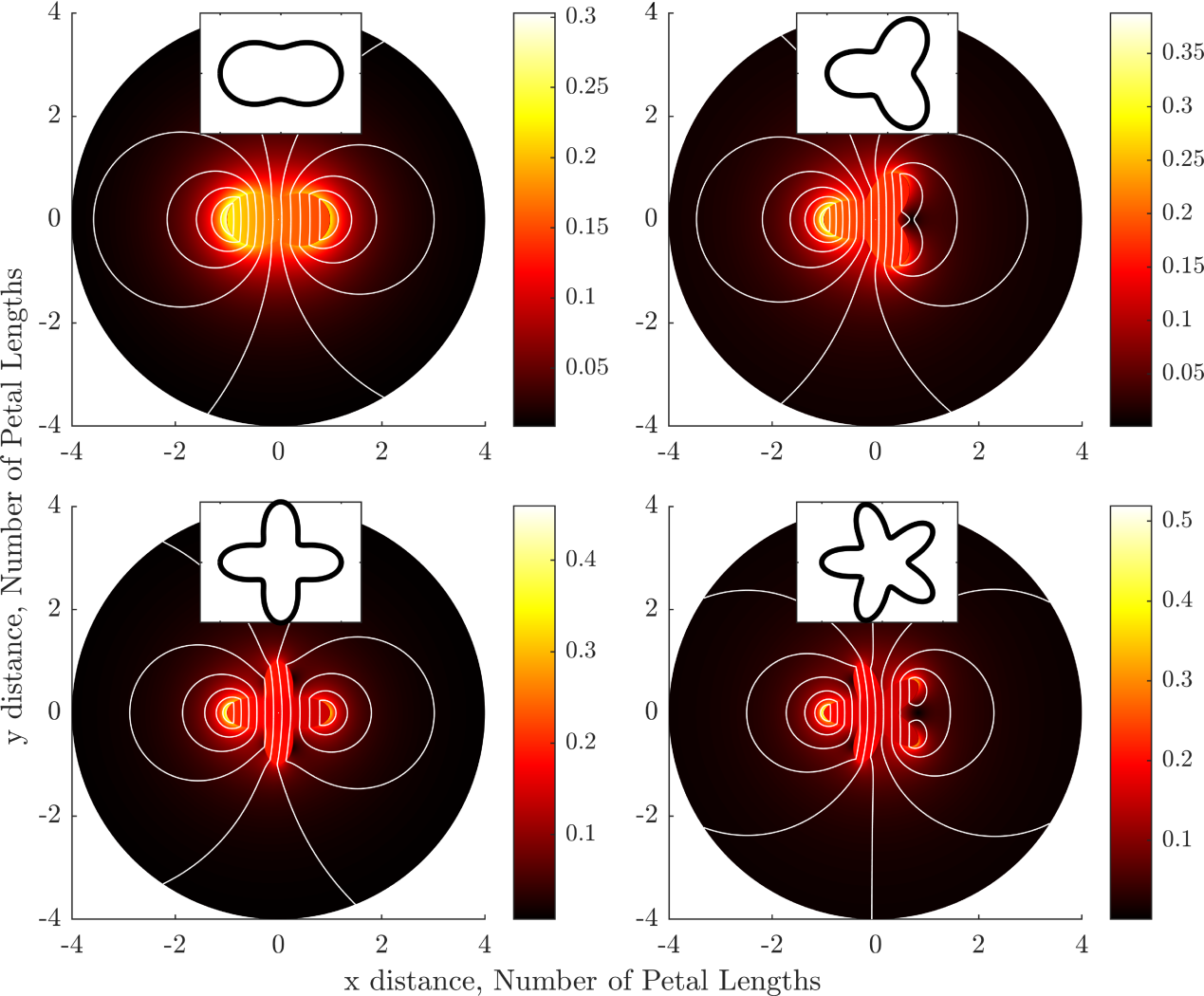}
    \caption{Perturbation electrical field magnitude $\|\mathbf{E}_P\|$ (colourmap) and potential $V_P$ (contours) for an uncharged flower polarising in the presence of a charged pollinator (external point charge).}
    \label{fig:bee-flower}
    \includegraphics[width=0.8\linewidth,trim=1 1 1 1,clip]{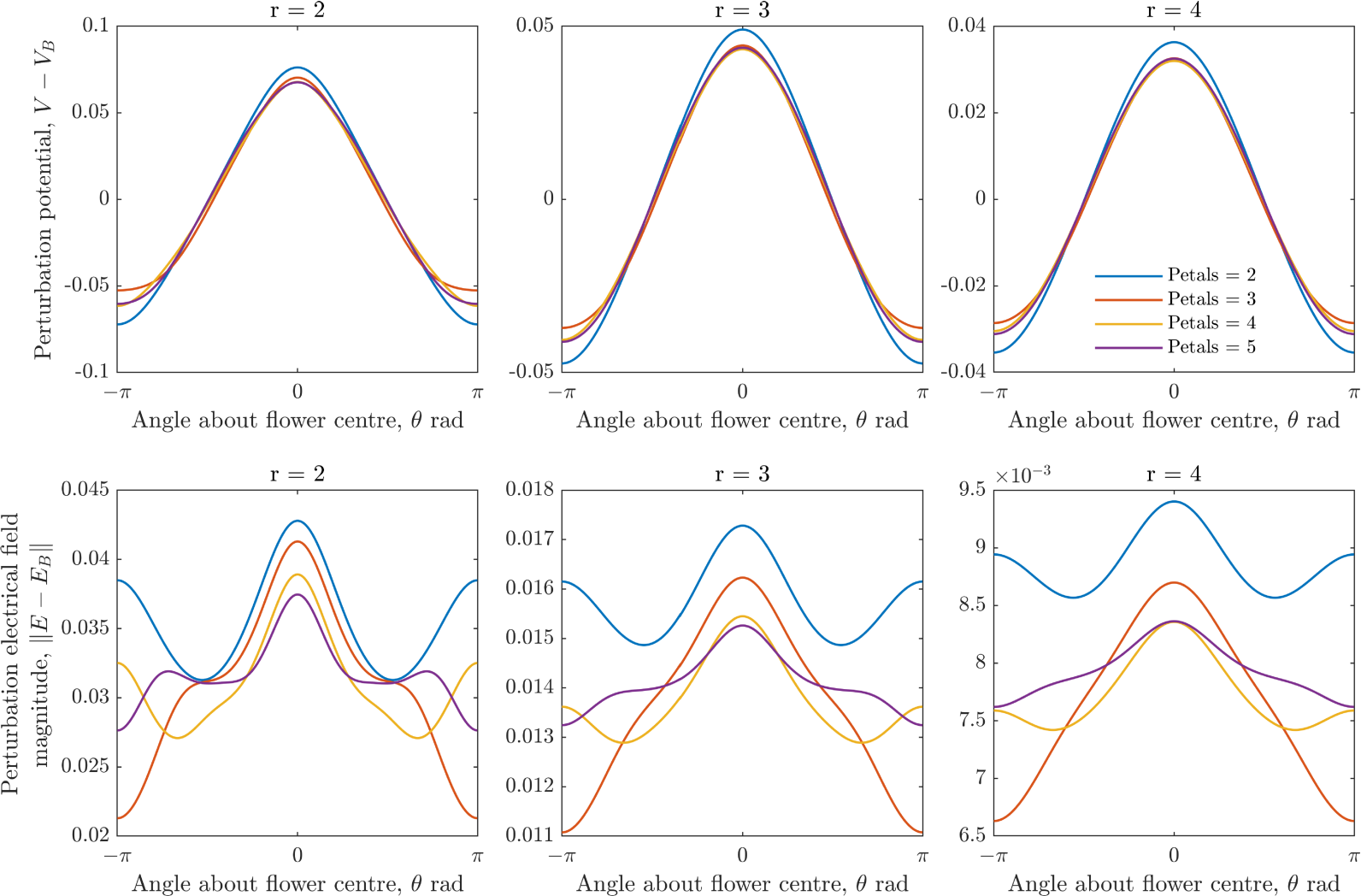}
    \caption{Perturbation electrical potential and field magnitude of flower polarisation in the presence of a pollinator (external point charge) at increasing radii from the flower centre, $r = 2, 3, 4$.}
    \label{fig:bee-flower-VE}
\end{figure}

Finally of note is the fact that, due to the nondimensionalisation, the only other parameter to vary here is the relative permittivity of the flower.
In Figure \ref{fig:uni-flower-perm}, we compare the flower perturbation field and potential of a three petal flower for different permittivities $\epsilon_R = 2, 10, 50, 250$ at distance $r = 2$.
For small values, any variation in the relative permittivity has a larger effect on the perturbation field and potential.
Yet, for values of $\epsilon_R \sim \mathcal{O}(10)$, less variation is seen as the field tends to a limit. 
Therefore, our results for $\epsilon_R = 20$ reasonably represent the characteristics of biological scenarios and where $\epsilon_R$ may vary between species.

\subsection{Pollinator-flower interactions}
\label{subsec:p-f-interactions}
\subsubsection{A pollinator's presence}
\label{subsubsec:p-f-bees}
The next scenario we consider is that of a charged pollinator, say a bee, in the presence of an uncharged flower.
With the bee's electrical field acting as the source, the flower is considered as polarising in response.
The perturbation field is now given by the bee's electrical field in the presence of a flower minus its field without a flower. 
Treating the bee as a point source external to the flower, we utilise the set-up in section~\ref{subsec:bee}.
In addition to the flower's relative permittivity ($\epsilon_R = 20$), the point source's location is an additional parameter that can be varied (the effect of which will be investigated in due course).
In the proceeding examples, the point source is located at $x = -5$, $y = 0$, unless stated otherwise.

The perturbation electrical field (colourmap) and potential (contours) are shown in Figure \ref{fig:bee-flower} both internally and externally for flower shapes of two to five petals. 
Overall, the results are largely comparable to those of the uniform electrical field in Figure \ref{fig:uni-flower}.
Notable differences include smaller electrical field magnitudes by a factor of 4 in each case (this is due in part to the different physical underpinnings of the two scenarios and therefore nondimensionalisations).
The vertical symmetry is also now lost, with mild discrepancy for the even flower shapes. 
Otherwise, the trends are of remarkable similarity to the uniform case, as are the forms of the potential and electrical field magnitude curves for the equivalent radial measures in Figure \ref{fig:bee-flower-VE}.

\begin{figure}
    \centering
    \includegraphics[width=0.9\linewidth]{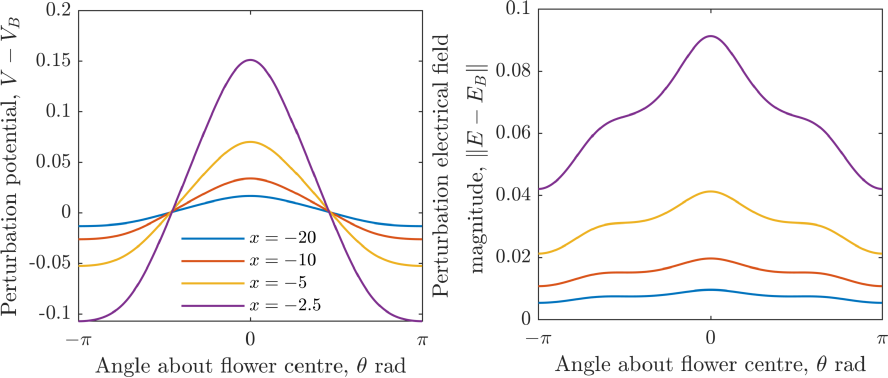}
    \caption{Three-petal flower polarisation due to an external charge at $x = -20, -10, -5, -2.5$. Left: Perturbation electrical potential; Right: Perturbation electrical field magnitude. Both are at $r = 2$.}
    \label{fig:bee-flower-x}
\end{figure}

In Figure \ref{fig:bee-flower-x} we consider the influence of the external charge location varying $x = -20, \ -10, \ -5, \ -2.5$.
The floral perturbation potential and field magnitude increase non-linearly as a pollinator approaches a flower.
Hence, the strength of the information the pollinator receives increases, as does the electrical floral signal in all directions.

\begin{figure}
    \centering
    \includegraphics[width=0.9\linewidth,trim=1 1 1 1,clip]{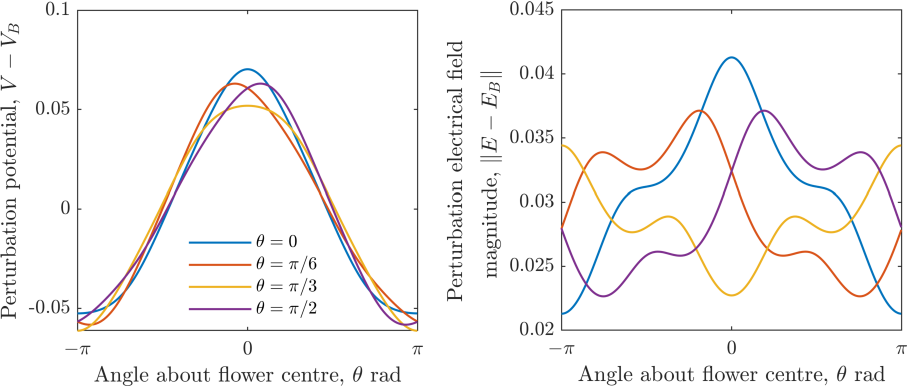}
    \caption{Three-petal flower polarisation due to an external charge, with the flower rotated by $\theta = 0, \pi/6, \pi/3, \pi/2$ (between axes of rotation symmetry). Left: Perturbation electrical potential; Right: Perturbation electrical field magnitude. Both are at $r = 2$.}
    \label{fig:bee-flower-rad}
\end{figure}

Considering the pollinator's location further, in Figure \ref{fig:bee-flower-rad}, we compare the same metrics again for a flower rotated by an angle $\theta = 0, \pi/6, \pi/3, \pi/2$ with a point charge located at $x = -5, \ y = 0$.
This simulates the effect of a pollinator's approach to different flower orientations.
The results show that by approaching a petal tip (blue curve) the perturbation potential and electrical fields are strongest, largely due to the pollinator being closer to the flower.
Conversely, when approaching the ``trough'' between two petals, the potential is much smaller and, due to the floral geometry, produces a minimum in the electrical field.
While the overall trend holds, e.g. a stronger electrical field near the petal tips that decreases with the flower geometry, the approach of a pollinator can greatly affect the perturbation field and the electrical signal that a flower propagates. 
For example, smaller overall values occur when $\theta = \pi/2$. 

\begin{figure}
    \centering
    \includegraphics[width=0.85\linewidth,trim=1 1 1 1,clip]{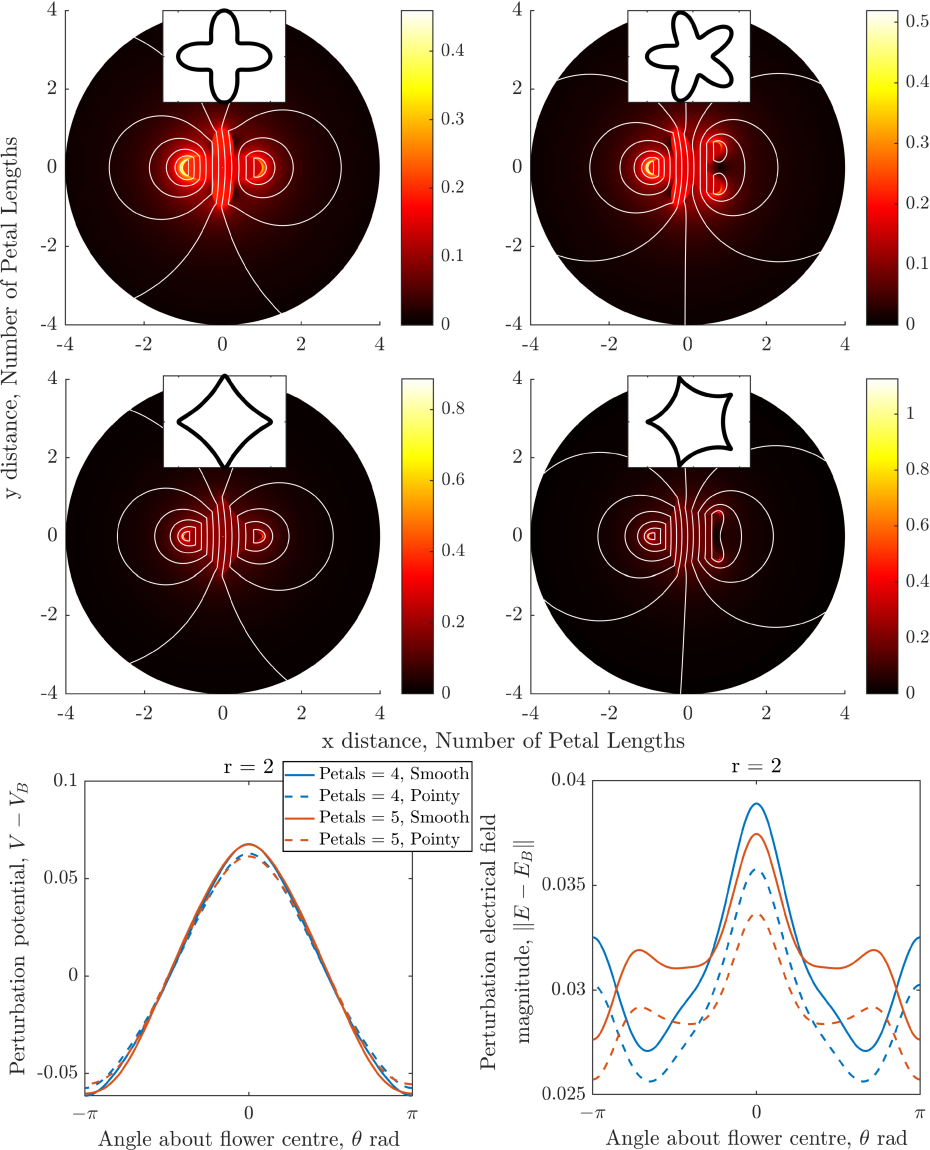}
    \caption{Perturbation electrical field magnitude $\|\mathbf{E}_P\|$ (colourmap) and potential $V_P$ (contours) for an uncharged flower polarising in the presence of a charged pollinator (external point charge). Four flower shapes are shown.
    Top row: flowers with four and five smooth petals.
    Middle row: flowers with four and five pointed petals.
    Bottom row: Perturbation electrical potential and field magnitude, each at $r = 2$.}
    \label{fig:bee-flower-shape}
\end{figure}

Finally, in Figure \ref{fig:bee-flower-shape}, the perturbation electrical field (colourmap) and potential (contours) of four flower shapes are shown to compare smooth curved petals and sharp pointed petals. 
The radial measures of each are shown in the bottom row.
The perturbation electrical fields of the pointed flowers display higher values close to the flower boundary due to the sharp gradients here. 
Once again, the perturbation potential is reasonably consistent across flower shapes.
However, at two petal lengths from the flower centre, the smooth petals produce a higher field strength. 
The reasoning is that while the pointed flowers display a strong peak, the charge is more evenly distributed in the smooth petal case resulting in higher field strengths further away. 

\subsubsection{The electrostatic presence of pollen}

In theory, as a charged pollinator approaches a flower, both its petals and pollen polarise in the pollinator's presence.
How does pollen alter the surrounding electrical field and the overall floral signal?
Since the pollinator remains static in our current problem, we treat the pollen as a fixed point charge at the centre of the flower. 
By polarisation, its charge is a function of the pollinator's charge and of opposite sign.
In the analysis below, we treat the pollen as equally and oppositely charged to the pollinator. 
This is reasonable since the results scale linearly with the ratio of pollinator-pollen charge and illustrate the general effect.

\begin{figure}[htb!]
    \centering
    \includegraphics[width=0.84\linewidth]{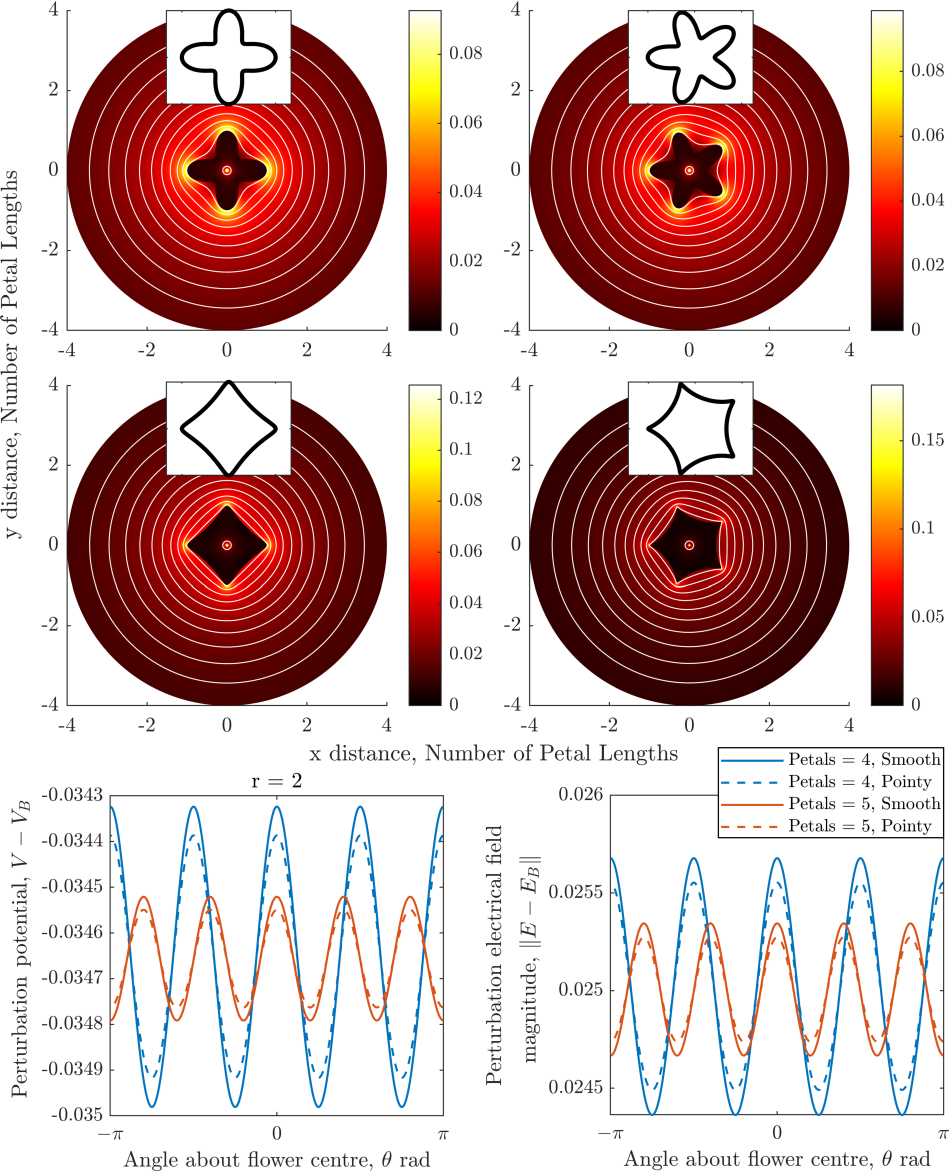}
    \caption{Perturbation electrical field magnitude $\|\mathbf{E}_P\|$ (colourmap) and potential $V_P$ (contours) for an uncharged flower polarising in the presence of a charged pollinator (external point charge) and charged pollen (internal point charge). Four flower shapes are shown.
    Top row: flowers with four and five smooth petals.
    Middle row: flowers with four and five pointed petals.
    Bottom row: Perturbation electrical potential and field magnitude, each at $r = 2$.}
    \label{fig:bee-pollen-flower}
\end{figure}

In Figure \ref{fig:bee-pollen-flower} the perturbation electrical field (colourmap) and potential (contours) are shown for the same flower shapes as in Figure \ref{fig:bee-flower-shape}.
The source field consists of a point charge external to the flower ($x = -5$ and $y = 0$), and an internal charge ($x = 0$ and $y = 0$).
The perturbation field is obtained by subtracting the electrical field without pollen from the field with pollen.
The radial measures for perturbation potential and field magnitude are shown in the bottom row. 
These results show radial symmetry with the petals producing strong electrical fields at their tips and symmetrical contour lines.
The contour lines (perturbation potential), resemble the floral shape and the number of petals, differing from previous perturbation potential plots.
With increasing distance from the flower, they become circular.

Investigating further, we see that the potential and electrical field magnitude curves at $r = 2$ show peak potential and electrical field values aligning with the petal tips, and are equal for each petal.
The magnitude of these peaks decreases with the number of petals.
The overall structures of both results are similar to the potential and electrical field of only charged pollen and a flower (i.e. no external charge); yet, by considering the presence of a pollinator, we see that the situation is more ecologically relevant.
Since the results scale linearly with the pollen-pollinator charge ratio, the flower signal increases with the pollen charge which represents the case of higher amounts of pollen and thus stronger electrical fields overall. 

\subsection{Electro-floral communication: Multiple arthropods}
In this section we consider how the presence of multiple arthropods may electrically interact with the flower and thus enable or prevent the detection of both the arthropods and the flower.

\subsubsection{Two bees or not two bees?}

\begin{figure}
    \centering
    \includegraphics[width=0.8\linewidth]{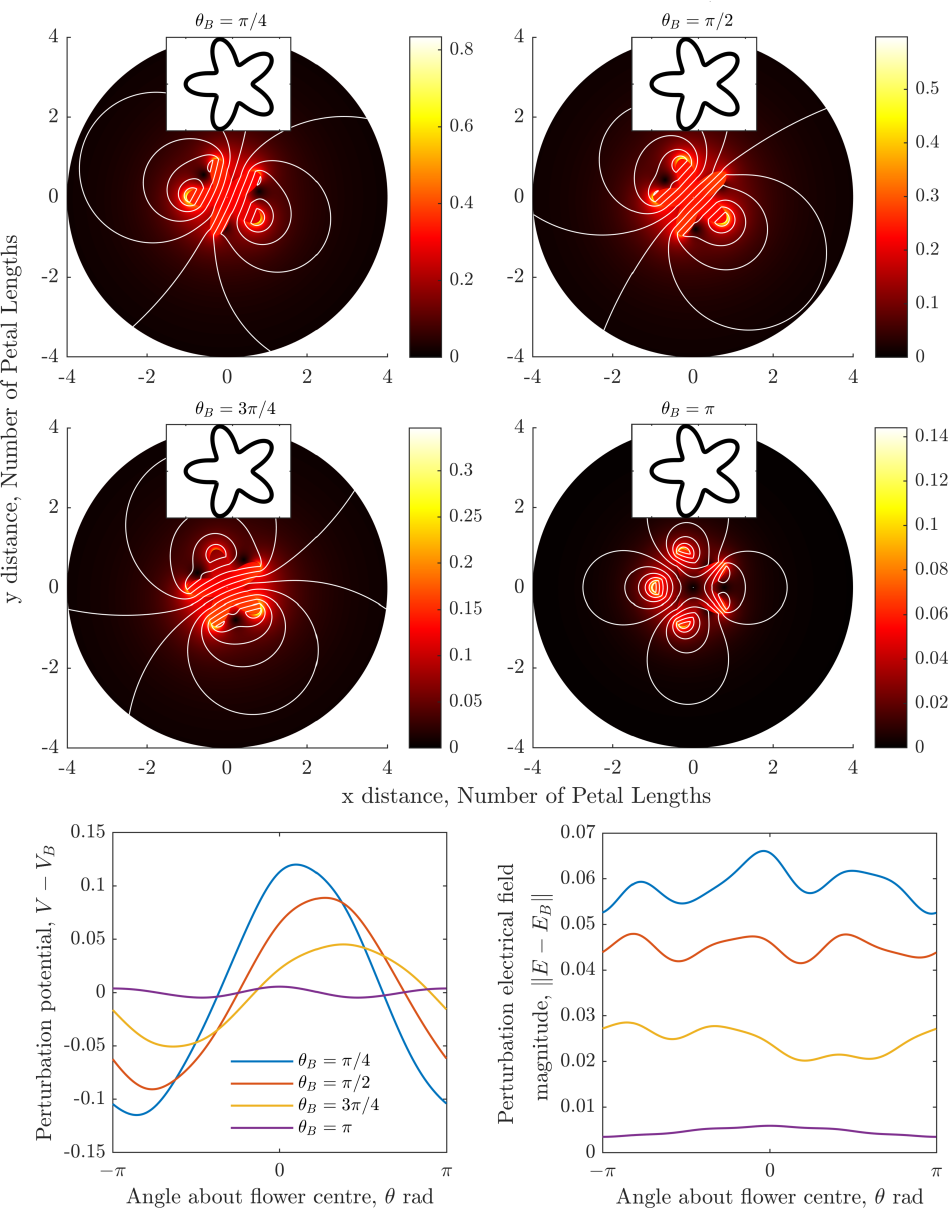}
    \caption{Top four plots: Perturbation electrical field magnitude $\|\mathbf{E}_P\|$ (colourmap) and potential $V_P$ (contours) for an uncharged five-petal flower polarising in the presence of two charged pollinators (external point charges at $(-5,0)$ and $(-5\cos(\theta_B), \ 5 \sin(\theta_B))$, for $\theta_B = \pi/4, \ \pi/2, \ 3\pi/4,$ and $\pi$).
    Bottom row: Perturbation electrical potential and field magnitude. Both are at $r = 2$.}
    \label{fig:two-bee-5}
\end{figure}

Consider the scenario where two charged pollinators approach the same flower.
Assuming both pollinators are of equal charge and approach from different directions,
the question here is whether the flower's polarisation communicates their presence to each other.
To this end, take the perturbation field to be that of the flower and pollinators compared to the scenario without the flower.
How does the flower change the local electrical information?

The first arthropod takes position at $(-5,0)$, and we consider the position of the second to vary according to $(-5\cos(\theta_B), \ 5 \sin(\theta_B))$, for $\theta_B = \pi/4, \ \pi/2, \ 3\pi/4,$ and $\pi$.
In Figure \ref{fig:two-bee-5} the perturbation electrical field (colourmap) and potential (contours) are once again shown for flowers with five petals.
Furthermore, radial measures of the perturbation potential and field magnitude are shown.

The presence of the flower and extra arthropod reorients the local perturbation field in comparison to Figure \ref{fig:bee-flower}.
As expected, the field values obtained are higher here, with multiple strong peaks appearing. 
The presence of the flower thus strongly influences the interaction of multiple pollinators and the flower can electrically communicate the presence of several pollinators through the change in local field geometry and strength.
The potential and field strength magnitudes fall as the second arthropod moves further around the flower. 
When co-located on the same side of the flower, the two arthropods serve to increase the flower's polarisation, enhancing its electrical field contribution and thus its perturbation field.
However, as the second arthropod moves around the flower, each pollinator's polarisation effect begins to cancel out, greatly reducing the floral signal here.
Thus, at each location around the flower, the perturbation field takes a distinct form that can act as a source of information.
Most interestingly, when the pollinators are on opposite sides, the diminishing of the perturbation field and potential to zero is a clear indicator of the presence of multiple pollinators.

The perturbation potential form is largely similar to that of Figure \ref{fig:bee-flower} for a single pollinator, yet slightly perturbed such that the peak is between the two pollinators.
Moreover, the perturbation field displays a very different form that lacks symmetry and reveals the flower's morphology (four petal results compared to five) through the arrangement of the peaks and troughs.
Overall, the flower's polarisation can distinctly detail the presence of multiple charged arthropods at a distance. 

\subsubsection{Electrical subterfuge}
\begin{figure}
    \centering
    \includegraphics[width=0.7\linewidth]{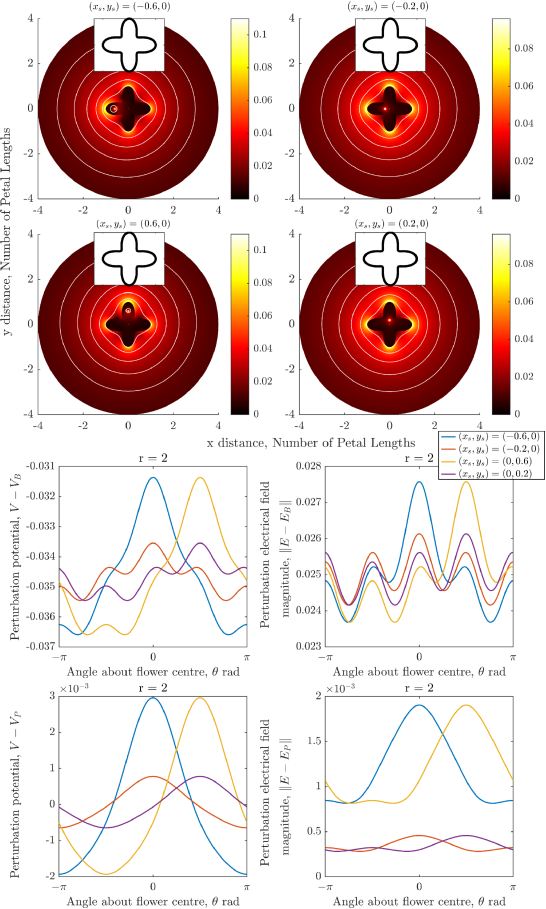}
    \caption{Perturbation electrical field magnitude (colourmap) and potential (contours) for flower polarisation in the presence of a charged pollinator (external charge, $x = -5, y = 0$) and a spider (internal charge, $(x_i, y_i) = (-0.6,0), \ (-0.2,0), \ (0,0.2), \ (0,0.6)$). 
    The colourmap/contour plots and third row compare the pollinator-spider scenario to that without a spider. 
    The bottom row compares the pollinator-spider scenario to the pollinator pollen scenario, e.g. a point charge at $(0,0)$. }
    \label{fig:fly-spider}
\end{figure}

Our final scenario for consideration is that of a predator-prey interaction.
This case may be interpreted one of two ways, 1) a predator (such as a crab spider) is hiding within the flower's petals, awaiting the arrival of a charged pollinator \cite{heiling2006picking, huey2017foraging}; or, 2) a charged predator (such as a wasp) is approaching a flower upon which a potential prey is located.

This scenario is similar to that of the pollen problem above, whereby the flower dwelling arthropod is treated as a point charge on a petal. Having considered the role of geometry in the similar pollen problem (Figure \ref{fig:bee-pollen-flower}), we now consider the role of the leaf-dweller's location.
Note that again due to the static nature of the problem, we treat the leaf-dweller as a fixed point charge of equal, opposite charge. 

Examining a four petal flower, we see that the perturbation field is given by comparing the electrical field with and without the flower-dweller present.
This enables us to assess how ``far'' the electrical presence of the dweller may be detected.
The exterior point charge is located at $(x_e, y_e) = (-5,0)$, and the following four internal positions of the internal charge are considered: $(x_i, y_i) = (-0.6,0), \ (-0.2,0), \ (0,0.2), \ (0,0.6)$.

When the internal point charge is closer to the petal boundary, $(-0.6,0)$ and $(0,0.6)$, the surface and contour plots show stronger perturbation electrical field magnitudes and greater deformation in the potential contours near the flower.
The radial metrics displayed in the third row (the comparison with and without internal charge) show this effect at a two petal lengths away from the flower centre.
However, away from the peak, all other perturbation field values are of a similar order.
Thus, overall, when a charge is closer to the flower centre it leads to a less distinct electrical field perturbation.
In comparison to the pollen scenario (bottom row), the difference between the two cases is very small indeed.
Thus, if a flower-dwelling arthropod seeks to cloak/hide itself on the flower, it is best to remain closer to the centre.
Indeed, the final row of plots shows that a spider on a flower can produce a perturbation field that is very similar to an equivalent pollen scenario (see \cref{subsubsec:p-f-bees}.
This enhances the possibility of a predator's electrical subterfuge.

Finally, we note that, similar to the pollen case, the magnitude of the internal/external charge ratio affects the perturbation field linearly. 
Thus, smaller internal charge values will produce smaller perturbation fields. 
Since the flower-dweller is expected to behave inductively and thus electrically equalise with the substrate, we expect a smaller relative charge value and hence greater cloaking (this is in contrast to the pollen scenario where we expect the field strength to be enhanced).

\section{Comparison to 3D - Pollinator-flower interaction}\label{sec:3D}
Following our examination of several ecologically relevant arthropod-flower electrostatic interactions, we now briefly compare the results of the AAA-LS method with those produced in COMSOL 6.1 using a 3D FEM model.
Currently, there is little empirical data on the shape and strength of floral electrical fields, and indeed electrical fields in general, due to a lack of technology, equipment or methodology for spatially measuring and visualising electrical fields in the real world. Therefore, we are unable to compare our results to empirical real-world data.

However, currently one of the best tools for visualising electrical fields and developing a spatial understanding of them is FEM. Therefore, to evaluate our methods, we seek a quantitative and qualitative comparison to FEM as the current state-of-the-art methodology. Since the main concern of the paper is to understand the role of floral morphology in the formation of floral signals, it is in this regard that we evaluate the comparison.

In the FEM model, the flower is treated as a dielectric, with a relative permittivity of 20, as in the AAA-LS analysis above. 
We only consider a three petal case and present nondimensional results.

\begin{figure}
    \centering
    \includegraphics[width=\linewidth,trim=1 1 1 1,clip]{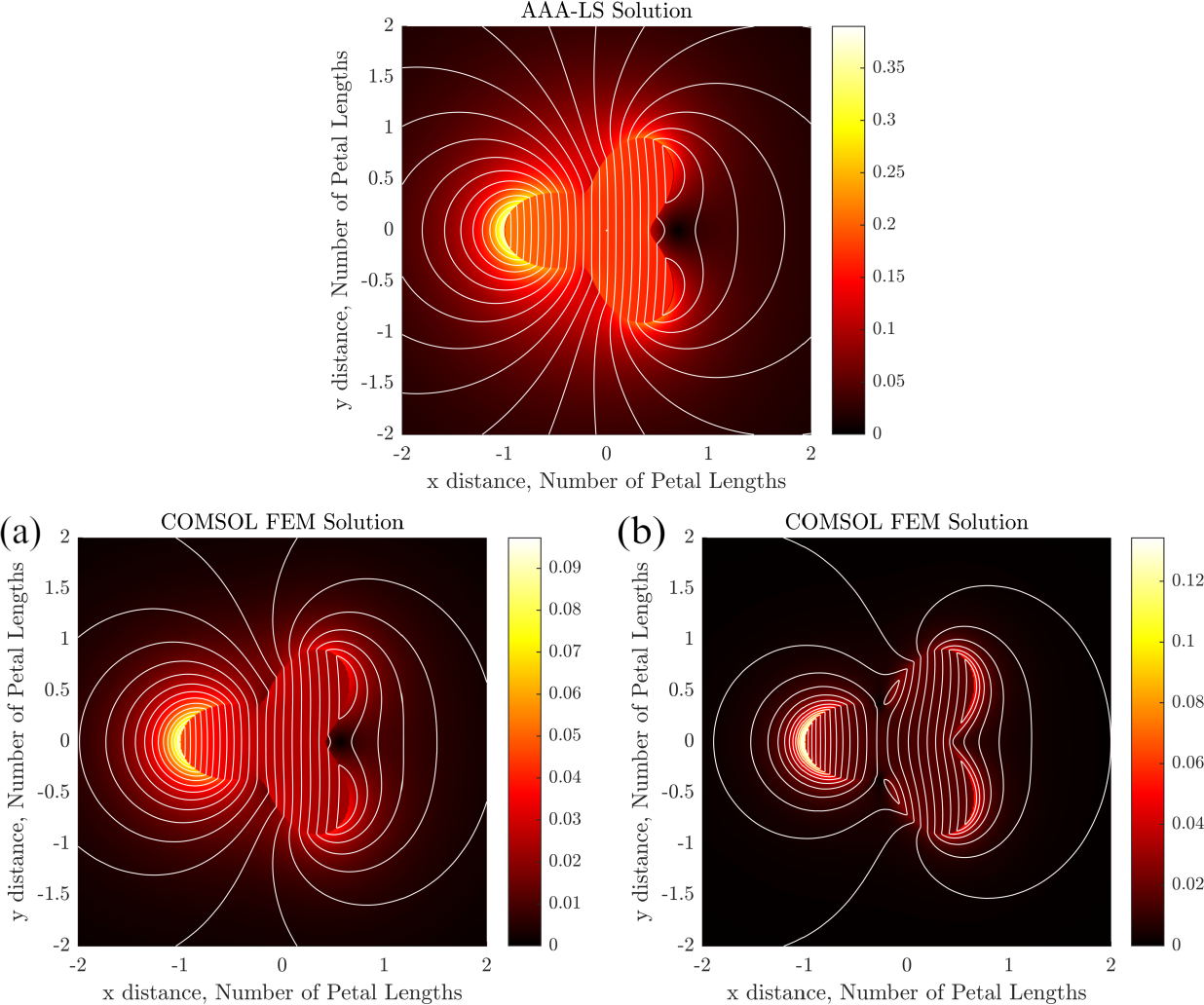}
    \caption{AAA-LS (top row) and FEM (bottom row) results for the perturbation electrical field magnitude $\|\mathbf{E}_P\|$ (colourmap) and potential $V_P$ (contours) for an uncharged flower polarising in the presence of a charged pollinator (external point charge). (a) A flower of depth 1 petal length, (b) A flower of depth 0.1 petal lengths.}
    \label{fig:3D-bee-flower}
\end{figure}

Consider the scenario presented in section ~\ref{subsubsec:p-f-bees} and the three petal flower results in Figure \ref{fig:bee-flower}.
The same analysis is conducted on a 3D flower shape using an FEM model, as shown in Figure \ref{fig:3D-bee-flower}. 
The top row of Figure \ref{fig:3D-bee-flower} shows the AAA-LS result (which matches that of Figure \ref{fig:bee-flower}).
The FEM flower extends in the $z$-direction (i.e. orthogonal) to the $x-y$ plane.
The bottom row shows the results for the FEM model for two cases of this 3D flower with a thickness of (a) 1 petal length, $L$, and (b) 0.1 petal lengths, $0.1 L$.
For the FEM model, the external point charge is produced using a sphere of radius 0.005 petal lengths, with a uniform surface charge producing an electrical source equivalent to the point charge case.
The presented results are taken from a slice through the flower's midpoint ($z = 0$) parallel to the $x-y$ plane.

Overall, there is notable qualitative agreement between the AAA-LS and FEM results.
When the flower is of larger thickness, the results match closely in terms of field variation and the general shape of the contours. 
Indeed, for the petal thickness of $L$, the somewhat semicircular contours at the petal tips and the 
deflection in the contour at the back of the flower are captured.
This flower however is somewhat unphysical since real-world flowers are often much thinner. 
The case for the flower of depth $0.1 L$ shows greater deviation for the AAA-LS results, yet near the petal tips, reasonable qualitative agreement is seen, with greater agreement in the far-field.

This comparison gives confidence that some of the salient 3D features are captured by the AAA-LS results, and thus show the power of the 2D analysis for examining these biological cases broadly. 
In many of the applications above, the main interest centres around the electrical field at distances of several petal lengths away from the flower, an area in which the AAA-LS method shows good qualitative agreement.

\section{Discussion and conclusions}
\label{sec:conc}
We present a novel  method for calculating 2D electrostatic interactions that arise between different electrical field sources and polarisable, dielectric objects.
Our work extends the AAA-LS method of \cite{trefethen2020numerical,costa2023aaa} to solve Laplace's equation for two-domain problems.
This method is able to compute the electrostatic potential and resulting electrical field for problems with multiple point sources/sinks (internal or external to the polarisable material) and a uniform background field which cause an object with a fixed boundary to polarise.

Our motivation for developing this method came from studying the role of floral geometry in flower-arthropod interactions.
Throughout the analysis of section~\ref{sec:results}, we show that the presence of the flower can produce distinct electrical ``signatures'' or perturbations within the electrical field revealing information about the flower's shape, the quantity of pollen available or the possible presence of a predator at distance from the flower. 

\subsection{Comments on the two-domain AAA-LS algorithm}
While the AAA algorithm and the so called ``lightning least-squares'' method have been in development for over five years \cite{nakatsukasa2023first}, the two-domain extension presented in this work of the combined AAA-LS method is new. The added mathematical complexity lies in the boundary conditions: instead of a single boundary condition relating the harmonic function $\phi$ to a given function $h$, there now exists a coupled pair of boundary conditions involving the two unknowns $V_1$ and $V_2$. Dealing with this added complexity by considering the boundary conditions of the combined quantity $V_1-V_2$ still uses the fundamentals from the original AAA-LS method; however, now quantities such as the list of poles and basis vector matrix are ``twice as large'' to account for the finding of two harmonic functions $V_1$, $V_2$ across two domains. 

Two-domain problems, also known as two-phase or ``Muskat'' problems in the context of fluid dynamics, are  difficult to solve both analytically and numerically \cite{crowdy2006exact}. The simple, fast, and accurate two-domain AAA-LS method developed in this work is thus applicable to a variety of other problems, including those in electro- and magnetostatics, such as noted in \cite{costa2023aaa}; fluid dynamics, such as the two-phase Hele-Shaw cell \cite{muskat1934two,howison2000note,zukowski2024breakthrough} and vortex dynamics \cite{wu1984steady}; phase-change scenarios such as the melting, freezing and dissolution of an interface \cite{goldstein1978effect,ladd2020}; and in biology, specifically in collective behaviours, such as the huddling of penguins in a cold wind \cite{harris2023penguin}.

\subsection{Comments on the biological and ecological relevance of results}
It has been hypothesised that the mutual electrostatic interaction between plants and arthropods is an additional sensory cue by which decisions can be made \cite{clarke2013detection, england2021ecology}.
Indeed, increasing empirical effort and evidence is beginning to shine a light on the nuances of this sensory modality and its wider role in arthropod ecology.

Our exploration of flower morphology and arthropod interaction makes several contributions to the wider biological discussion.
First, our results are presented in nondimensional form and have therefore been discussed in terms of ``petal lengths'' and scaled by the background field strength.
In nature, the lengths of flower petals can vary dramatically from a few millimeters to several centimeters. 
As such, larger flowers create a larger perturbation in the environmental field providing a detectable ``signal'' from much farther away.
For example, in section~\ref{subsec:bee}, we consider an hypothetical bee to be five petal lengths from the flower, resulting in a polarisation interaction and reciprocated electrical signal from the flower.
The effects of the flower's presence are seen to persist up to four petal lengths away (and, indeed, beyond this is many cases). 
Thus, for a flower on the millimeter scale the effects remain relatively localised, while larger, centimeter scales can lead to significant electrical perturbations at far greater distances.
All of this is relative to the background field strength, scaling linearly, and again shows the potential for stronger fields to also lead to far more significant perturbations at a distance.
For example, bees have been show to carry charge $\mathcal{O}(100) \ pC$, which can produce biologically sensitive fields within the flower at this scale.
To re-scale the electrical fields and potentials to reflect this case, each result needs to be multiplied by $10^{-7}/2\pi \epsilon_0 \approx 1800$.
Hence, the perturbation electrical potentials are $ \mathcal{O}(50)$ to $ \mathcal{O}(100) \ V$ and the perturbation fields are $ \mathcal{O}(50) \ FC^{-1}$ (depending on the flower shape).
Such forces are strong enough to deflect arthropod mechanosensors \cite{rpalmer2021analysis, palmer2022mechanics}, highlighting the potential sensory significance of these flower-arthropod interactions.

Second, in each of our analyses the shape of the flower leads to strong qualitative and quantitative changes in the electrical field shape and strength. 
Thus, considering the broader variation in flower morphology \cite{byng2018phylogeny}, our results indicate that significant morphological variation can lead to even more distinct signals and convey potentially unique information about the flower according to its shape.
Indeed, several aspects of flower morphology and heterogeneity are not captured by our analysis (e.g. large 3D variations, plant sex organs \cite{byng2018phylogeny}) such that we anticipate the presence of these features to produce even stronger perturbations and interactions. 

Regarding the suggested symmetry of electrical fields relative to floral morphology and pattern detection, it is well known that floral symmetry is a significant factor within pollinator decision-making \cite{giurfa1999floral, endress2001evolution, citerne2010evolution}. From visual cues of symmetry, a pollinator is able to assess the health of a flower, the abundance of available resource (nectar or pollen),  and the possible presence of a predator \cite{heiling2006picking, huey2017foraging}. We hypothesise that this co-evolved sensitivity to vertical, horizontal and radial symmetry will also be electrically discernible and important to an arthropod, providing further relevant information about the scenarios listed above.

Two recent studies are informative in this regard. A theoretical study \cite{palmer2024sensing} shows the feasibility and possibility of shape detection using electrical fields via insect-like mechanosensors and for biologically relevant parameters. Furthermore, an empirical study \cite{woodburn2024electrostatic} reveals how flowers act as electrical antenna, producing electrical fields and signals in response to a surrounding/nearby field, and showing the influence of their shape on the strength of the signals they produce. Coupled with the acute sensitivity of arthropod sensory organs (neuronal responses can be elicited for small deflections of mechanosensory hairs in the order of $10^{-2}$ to $10^{-4}$ radians \cite{sutton2016mechanosensory, palmer2022mechanics}), small spatial and temporal variations in electrical fields are expected to be detectable and interpretable, thus revealing the sensory importance of floral geometry in the variation and production of electrical floral signals. Further research is required in this area to evaluate and confirm these hypotheses and would provide valuable directions for empirical exploration.

Third, our results indicate the additional conveyance of information to an external arthropod due to the presence of pollen or predators on the flower.
While modest in nondimensional size, as seen above, when scaled to real-world values, each case presents a relatively strong perturbation that once again may convey ecologically relevant information (such as the amount of pollen or a hidden predator) that would be a vital additional consideration for an arthropod.
Regarding the predator-prey interaction, the perturbation fields shown here are the smallest of all the results.
Indeed, it is in the predator's interest to have a minimal effect on the electrical field so as to remain undetectable.
Thus, the interpretation of the predator-prey results may be best viewed through the lens of the predator presenting a smaller charge to minimise its electrical presence.
The dynamics of such predator-prey interactions from an electrostatic perspective would form a fascinating follow-up study.

\subsection{Limitation of method and work}

While the 2D results match well for thick flowers, more biologically accurate geometries (in 3D) would require thinner petals. 
Furthermore, 3D effects are important in the true, biological setting. 
In general, the 2D results provide a qualitative intuition and understanding of the 3D and biological scenarios and help provide evidence of interesting directions for future work. 
Indeed, the speed and adaptability of the AAA-LS method enables the opportunity to analyse a vast number of biological scenarios and draw biologically relevant conclusions for further empirical and modelling investigation.

Finally, the flowers have been treated as dielectrics and the arthropods as point charges.
While petals and leaves generally present dielectric properties on their surfaces, conductivity can play a significant role in electrostatics plant problems.
For example, rainfall will increase surface conductivity and change plants' electrical properties. 
In addition, the point charge approximation is most appropriate when the petal length is much greater than the arthropod/pollen size. 
Hence, there is a limitation in the scale/size of both the arthropod and the flower that may be accurately considered.

\subsection{Future work}
There are several areas of work that would benefit from further theoretical and empirical research.

First, the two-domain AAA-LS method can be readily adapted to  fluid flow scenarios (e.g. potential flows) or magnetic fields, mostly requiring a change to the boundary and far-field conditions. 
Second, the method could be adapted to consider conductive media.
The underlying equations change from Laplacian to Poisson which adds considerable complexity. 
One solution is to consider an iterative method to account for reciprocal electrical interactions in the system. 
Third, moving from a two-domain problem to an $N$ domain problem or to a system with heterogeneous materials will allow for a broader range of physical scenarios to be modelled.  
Fourth, the system can be readily adapted and used to investigate time-dependent predator-prey and pollinator-flower dynamics. 
Namely, under the assumption that the electrical fields vary quasi-statically in comparison to the timescale of insect movement and sensation \cite{palmer2023analysis}, moving arthropods and dynamic floral responses can be readily incorporated. This is a potentially fruitful avenue of future work with recent results highlighting the role of time-varying electrostatics for predators and prey \cite{england2024prey, oreilly2024}.

Finally, systematic empirical and 3D modelling studies would further reveal the ecological relevance of floral signals.
A deeper understanding of environmental and ecological electrical interactions will add to our knowledge of plant-pollinator coevolution and enhance our understanding of the pervasive sensory modality of electroreception that may be of ubiquitous use throughout the natural world.

\appendix
\section{Exact solutions}
\subsection{Uniform field}\label{appendix1}

Two exact solutions for a flower in a uniform electrical field, \eqref{Meq:ilap}-\eqref{Meq:ufar}, are now given. First, for a circular flower $\gamma: r=1$, interior and exterior electric potentials are
\begin{gather}\label{exactunicircle}
    V_1=-\frac{2\tilde{\epsilon}}{1+\tilde{\epsilon}}r\cos\theta,\\
    V_2=-r\cos\theta+\frac{1-\tilde{\epsilon}}{1+\tilde{\epsilon}}\frac{\cos\theta}{r}.
\end{gather}

Second, for an elliptical flower, a procedure similar to that of \cite{crowdy2006exact} is used based on the Schwarz function $S(z)$, which is an analytic function in the neighborhood of $\gamma$ such that $S(z)=\bar{z}$ on $\gamma$. Let
\begin{equation}
    \E_1 = -\boldy{\nabla}V_1=\frac{1}{\epsilon_1}\boldy{\nabla}\phi_1, \;\;\;\;\; \E_2 = -\boldy{\nabla}V_2=\frac{1}{\epsilon_2}\boldy{\nabla}\phi_2,
\end{equation}
where
\begin{equation}\label{pot}
    \phi_1 = -\epsilon_1V_1,\;\;\;\;\phi_2 = -\epsilon_2V_2.
\end{equation}
This gives a problem similar to that considered by \cite{crowdy2006exact} for the time-dependent deformation of an elliptical inclusion in a two-phase Hele-Shaw strain flow.

The harmonic potentials $\phi_1$ and $\phi_2$ are real parts of the complex analytic functions $w_1(z)$ and $w_2(z)$, respectively. Following \cite{crowdy2006exact}, the boundary conditions \eqref{Meq:dBC} and \eqref{Meq:nBC} are used to obtain an expression for $w_2'=dw_2/dz$:
\begin{equation}\label{w2prime}
    w_2'(z) = \frac{1}{2}\Big((\tilde{\epsilon}+1)w_1'(z) +(\tilde{\epsilon}-1)\overline{w_1'(z)}S'(z)\Big),
\end{equation}
recalling that $\tilde{\epsilon}=\epsilon_2/\epsilon_1$. Note that use of $S(z)$ in \eqref{w2prime} provides an analytic continuation away from $\gamma$. Now assume that the interior complex potential has the form
\begin{equation}
    w_1(z) = Uz, \label{eq:w1}
\end{equation}
where $U$ is a (real) constant to be found. Note that \eqref{eq:w1} is a different choice than that made in \cite{crowdy2006exact}. Putting \eqref{eq:w1} in \eqref{w2prime}, integrating and setting the irrelevant constant of integration to zero, gives
\begin{equation}\label{eq:w2}
    w_2(z) = \frac{U}{2}\Big((\tilde{\epsilon}+1)z+(\tilde{\epsilon}-1)S(z)\Big).
\end{equation}
Recall that
\begin{equation}
    \E_2 = \frac{1}{\epsilon_2}\boldy{\nabla}\phi_2\rightarrow \boldy{\hat{x}}\text{ as }r\rightarrow\infty.
\end{equation}
Using complex notation and result (5) from \cite{trefethen2018series}, we obtain
\begin{equation}
    E_2 = \frac{1}{\epsilon_2}\nabla[\Real(w_2(z))] = \frac{1}{\epsilon_2}w_2'(z)=\frac{U}{2\epsilon_2}\Big((\tilde{\epsilon}+1)+(\tilde{\epsilon}-1)\overline{S'(z)}\Big),
\end{equation}
and therefore
\begin{equation}
    U = \frac{2\epsilon_2}{\Real\big[(\tilde{\epsilon}+1)+(\tilde{\epsilon}-1)\lim_{z\to\infty}\overline{S'(z)}\big]}. \label{Meq:Ulim}
\end{equation}

To find  $U$,  note that the Schwarz function $S(z)$ for the ellipse with semiaxes $a+b$ and $a-b$ along the $x-$ and $y-$axes with $a>b\ge 0$, is \cite{davis}
\begin{equation}\label{eq:ellschwarz}
    S(z)  = \frac{b}{a}z+\frac{a^2-b^2}{2ab}(z-\sqrt{z^2-4ab}),
\end{equation}
where the square root branch is taken such that $S(z)=a^2z^{-1}$ for $b=0$, the well-known Schwarz function for a circle of radius $a$ centred at the origin. From \eqref{eq:ellschwarz} $\lim_{z\to\infty}\overline{S'(z)}=b/a$ and so \eqref{Meq:Ulim} determines $U$.

To summarise,  the exact solution for an ellipse from \eqref{pot}, \eqref{eq:w1} and \eqref{eq:w2} is
\begin{gather}
    V_1 = -\frac{U}{\epsilon_1}\Real[z], \label{Meq: eV1}\\
    V_2 = -\frac{U}{2\epsilon_2}\Real\bigg[(\tilde{\epsilon}+1)z + (\tilde{\epsilon}-1)\bigg(\frac{b}{a}z+\frac{a^2-b^2}{2ab}(z-\sqrt{z^2-4ab})\bigg)\bigg], \label{Meq:eV2}
\end{gather}
where
\begin{equation}
    U = \frac{2\epsilon_2}{(\tilde{\epsilon}+1)+(\tilde{\epsilon}-1)\frac{b}{a}}. \label{Meq:U}
\end{equation}

As in \cite{crowdy2006exact} this method works because of the special nature of the Schwarz function of the ellipse having unique behaviour $S(z)\sim Cz$ as $z\to\infty$ for some constant $C$. Such a property does not extend to other flower shapes for which analytic continuation of the Schwarz function invariably gives rise to singularities in the form of branch cuts and poles which do not have natural physical interpretations.

\subsection{Arthropod exact solutions}\label{appendix2}

An exact solution for the problem  \eqref{Meq:ilap} - \eqref{Meq:nBC}, \eqref{Meq:spfar} (a bee or spider in the presence of a flower) can be found for a circular flower boundary. The problem is analogous to that of a point vortex in the presence of circular topography -- see \cite{hinds2016beach}. Defining $\alpha = (\tilde{\epsilon}-1)/(\tilde{\epsilon}+1)$ and $\beta = 1+\alpha$, when the point charge is inside the circle $\lvert z_1\rvert<1$, i.e. a spider sitting atop the flower, the  potentials are \cite{hinds2016beach}
\begin{gather}
    V_1 = \tilde{\epsilon}\big(\log\lvert z-z_1\rvert - \alpha\log\big\lvert 1-z\overline{z_1}\big\rvert\big), \\ 
    V_2 = \beta\log\lvert z-z_1\rvert - \alpha\log\lvert z\rvert.
\end{gather}
For $\lvert z_1\rvert>1$, i.e. a bee flying outside the flower, the potentials are \cite{hinds2016beach}
\begin{gather}
    V_1 = \beta\log\lvert z-z_1\rvert - \alpha\log\lvert z_1\rvert, \\ 
    V_2 = \log\lvert z - z_1\rvert + \alpha\log\big\lvert1-1/(z\overline{z_1})\big\rvert.
\end{gather}
Since the problem is linear, solutions can also be found for a swarm of arthropods near a circular flower; the solution \eqref{exactunicircle} for a uniform field can also be included. 

\section*{Acknowledgments}
We thank colleagues in the University of Bristol's `Electrical Ecology Lab' for helpful discussions, namely Prof. Daniel Robert and Dr Liam O'Reilly. We also thank the referees for their useful suggestions. Samuel J. Harris was supported by a UK Engineering and Physical Sciences Research
Council PhD studentship, grant numbers EP/N509577/1 and EP/T517793/1.

\bibliographystyle{unsrt}  
\bibliography{ABbib}  

\end{document}